\documentclass[aps, prb, onecolumn, reprint, groupedaddress]{revtex4-2}

\usepackage{hyperref}
\usepackage{amsmath, amssymb}
\usepackage{braket}
\usepackage{multirow}
\usepackage{vtable}
\usepackage{algorithmic}
\usepackage{mathtools}
\usepackage{amsfonts}
\usepackage{extarrows} 
\usepackage{graphicx}
\usepackage{color}
\usepackage{bm}
\usepackage{hyperref}
\usepackage{cases}
\usepackage{comment}
\renewcommand{\Im}{\operatorname{Im}}

\begin{document}
\title{A computational method for angle-resolved photoemission
  spectra from repeated-slab band structure calculations}

\author{Misa Nozaki}
\email{6d697361@gmail.com}
\affiliation{Graduate School of Science and Engineering, Chiba University, Yayoi-cho 1-33, Inage, Chiba 263-8522, Japan}
\author{Peter Kr\"uger }
\email{pkruger@chiba-u.jp}
\affiliation{Graduate School of Science and Engineering, Chiba University, Yayoi-cho 1-33, Inage, Chiba 263-8522, Japan}
\date{\today}

\begin{abstract}
  A versatile method for angle-resolved photoemission spectra (ARPES)
  calculations is reported within the one-step model of photoemission.
  The initial states are obtained from a repeated-slab calculation
  using the projector-augmented wave (PAW) method.
  ARPES final states are constructed by matching the repeated-slab
  eigenstates of positive energy with free electron states that satisfy
  the time-reversed low-energy electron diffraction
  boundary conditions. Nonphysical solutions of the
  matching equations, which do not respect the flux conservation,
  are discarded.
  The method is applied to surface-normal photoemission from graphene
  as a function of photon energy from threshold up to 100 eV.
  The results are compared with independently performed multiple
  scattering calculations and very good agreement is obtained,
  provided that the photoemission matrix elements are computed with
  all-electron waves reconstructed from the PAW pseudo-waves.
  However, if the pseudo-waves are used directly,
  the relative intensity between
  $\sigma$- and $\pi$-band emission is wrong by an order of magnitude.
  The graphene ARPES intensity has a strong photon energy dependence
  including resonances.
  The normal emission spectrum from the $\pi$-band shows a
  hitherto unreported, sharp resonance at a photon energy of 31~eV.
  The resonance is due to a 2$D$ interband transitions and highlights
 the importance of matrix element effects beyond the final state plane-wave
 approximation.
\end{abstract}

\maketitle

\section{Introduction}
Angle-resolved photoemission spectroscopy (ARPES)
is the major experimental method
for determining the electronic structure of surfaces and low-dimensional
systems, including thin crystalline films and oriented molecules~\cite{ARPES}.
For two-dimensional systems, the electronic band dispersion
$E({\bm k}_\parallel)$ can be fully resolved, where $E$ is the energy and
${\bm k}_\parallel$ is the 2$D$ crystal momentum.
From the ARPES intensity distribution, the wave functions
can in principle be reconstructed with the orbital tomography
method~\cite{puschnig}.
In recent years, tremendous progress has been
achieved in instrumentation and data acquisition~\cite{PMM}, but
this is not matched on the theoretical side,
where a reliable and flexible calculation method is still missing.
To date, accurate ARPES spectra are mostly
computed with the multiple scattering method, especially in the layered KKR
implementation~\cite{pendry, braun, ono21}.
While the layered KKR method is efficient and accurate for dense,
inorganic matter, applications to open structures such as organic
materials have been severely hampered by technical difficulties of
full-potential multiple scattering theory~\cite{gonisbutler,keisuke}.

The ground state electronic structure of most materials is well
described by density functional theory (DFT) in conjunction with
the projector-augmented wave (PAW) method for the solution of the Kohn-Sham
equations~\cite{blochlpaw}.
In this scheme, surfaces and two-dimensional systems are modeled in
the repeated-slab geometry (RSG), i.e. by using periodic boundary conditions
in all three dimensions, where adjacent slabs are
separated by a sufficient amount of vacuum.
While the RSG is being used extensively for the ground state properties of
low-dimensional materials,
photoemission spectra cannot be obtained directly in this approach,
because the periodic boundary conditions of the RSG are incompatible
with the photoemission final state, which must evolve into
a free electron wave in the asymptotic limit.
Therefore the wave functions obtained in RSG must be
matched in some way to the far-field solutions which respect the
proper boundary conditions.
Previously,
Kraskovskii et al.~\cite{Kra04,Kra21} have developed an ARPES method based on
the RSG and a linearized augmented plane-wave (LAPW) calculation.
The method is very accurate for inorganic surfaces with
small unit cells, but has not been applied to more complex structures,
possibly because of numerical limitations of the LAPW approach~\cite{Kra20}.
Kobayashi~\cite{Kob20} has proposed an ARPES method based on the RSG and
PAW method~\cite{kressejoubert}, and applied it to low energy spectra of a Bi surface.
The approach presented here is conceptually similar to that of
Kobayashi~\cite{Kob20},
but we use a different matching procedure~\cite{Ono18} and
we go beyond the pseudo-potential level of PAW,
by computing the ARPES matrix elements from PAW-reconstructed
all-electron wave functions.
Moreover, we apply our method to a two-dimensional material and
study the photon energy dependence of the spectra,
while Ref.~\cite{Kob20} was devoted to spin-resolved ARPES from a metallic surface at low photon energy.
Let us note that all the approaches above, including the present one,
use the one-step model of photoemission which is based on
Fermi's golden rule and a stationary state description of the photoemission
final state~\cite{pendry}.
ARPES may also be calculated using time-dependent wave-propagation
techniques~\cite{dauth,horseshoe}.
While such a dynamical approach is mandatory
for time-dependent effects observed in ultrafast pumb-probe
experiments~\cite{optica}, it appears as an unnecessarily heavy numerical
technique for the calculation of time-independent ARPES spectra.

In this paper, we present a theoretical method for ARPES of 2$D$ systems,
based on a PAW band structure calculation in RSG.
The photoemission final states are obtained by matching the band states above
the vacuum level, to free electron waves with the proper, time-reversed LEED
boundary conditions.
The basic idea of the present matching method was anticipated
by Ono and Kr{\"u}ger~\cite{Ono18}, who studied a one-dimensional toy model.
Here we formulate the general theory, develop the computational method
and apply it to the prototype 2$D$ material graphene.
We focus on the normal-emission ARPES, excited
with $p$-polarized light, as a function of photon energy from threshold
up to about 100~eV.
We find that the ARPES intensity has a strong photon energy dependence
including resonances, in agreement with recent experiments~\cite{horseshoe}.
We compare the results
with independently performed multiple-scattering calculations
and find very good agreement over the whole photon energy range,
validating the accuracy of the present approach. 
The popular final-state plane-wave approximation, in contrast, fails to
reproduce the overall energy dependence and misses the resonance effects.
We examine the possibility of replacing the all-electron wave functions
by PAW pseudo-waves in the calculation of photoemission matrix elements.
We conclude that pseudo-waves must not be used, because the intensity ratio
between bands of different orbital character (here C-2$s$ and C-2$p_z$) can be
wrong by one order of magnitude.
For normal emission from the graphene $\pi$-band,
we predict a sharp Fano-resonance at a photon energy of 31~eV,
which has, to the best of our knowledge not been reported before.
We show that the resonance is due to the coupling
of an evanescent $\sigma$-type 2$D$ band-state with the free-electron
continuum.

\section{General Method}
Unless stated otherwise, we use atomic units where $\hbar$=$m_e$=$e$=1.
We employ the one-step model of photoemission which is based on
stationary scattering theory.
The photoemission intensity is given by \cite{Brown80}
\begin{equation}
  I(\epsilon_{f}, \omega, \hat{\bm p}) \propto \frac{1}{\omega} \sum_{i} \left|
  \braket{\psi_f^{-}|{\bm \varepsilon}\cdot{\hat {\bm P}}|\psi_i}\right|^2
  \delta(\epsilon_i + \omega -\epsilon_f)
  \label{eq:fermi}
\end{equation}

Here, $\omega$ is the photon energy, $\bm \varepsilon$ is the light polarization vector, 
$\hat{\bm p}$ is the direction vector of emitted photoelectron
and $\hat {\bm P}$ is the electron momentum operator.
$\psi_i$ is the initial state with energy $\epsilon_{i}$.
$\psi_{f}^{-}$ is the photoemission final state with energy
$\epsilon_{f}$ where the subscript indicates the time-reversed
LEED boundary conditions~\cite{Brown80}.
For simplicity we disregard the photoelectron spin
and assume that there is no absorption.
Then the time-reversed LEED state for
a photoelectron with momentum ${\bm p}$
is the complex conjugate of the LEED state with reversed momentum, i.e.
\begin{equation}\label{eq:TR}
  \psi_f^{-} = \psi_{-{\bm p}}^{*}
\end{equation}

\begin{figure}
\centering
\begin{tabular}{ll}
 \multicolumn{1}{c}{\includegraphics[bb=0 0 330 349, width=0.5\linewidth]{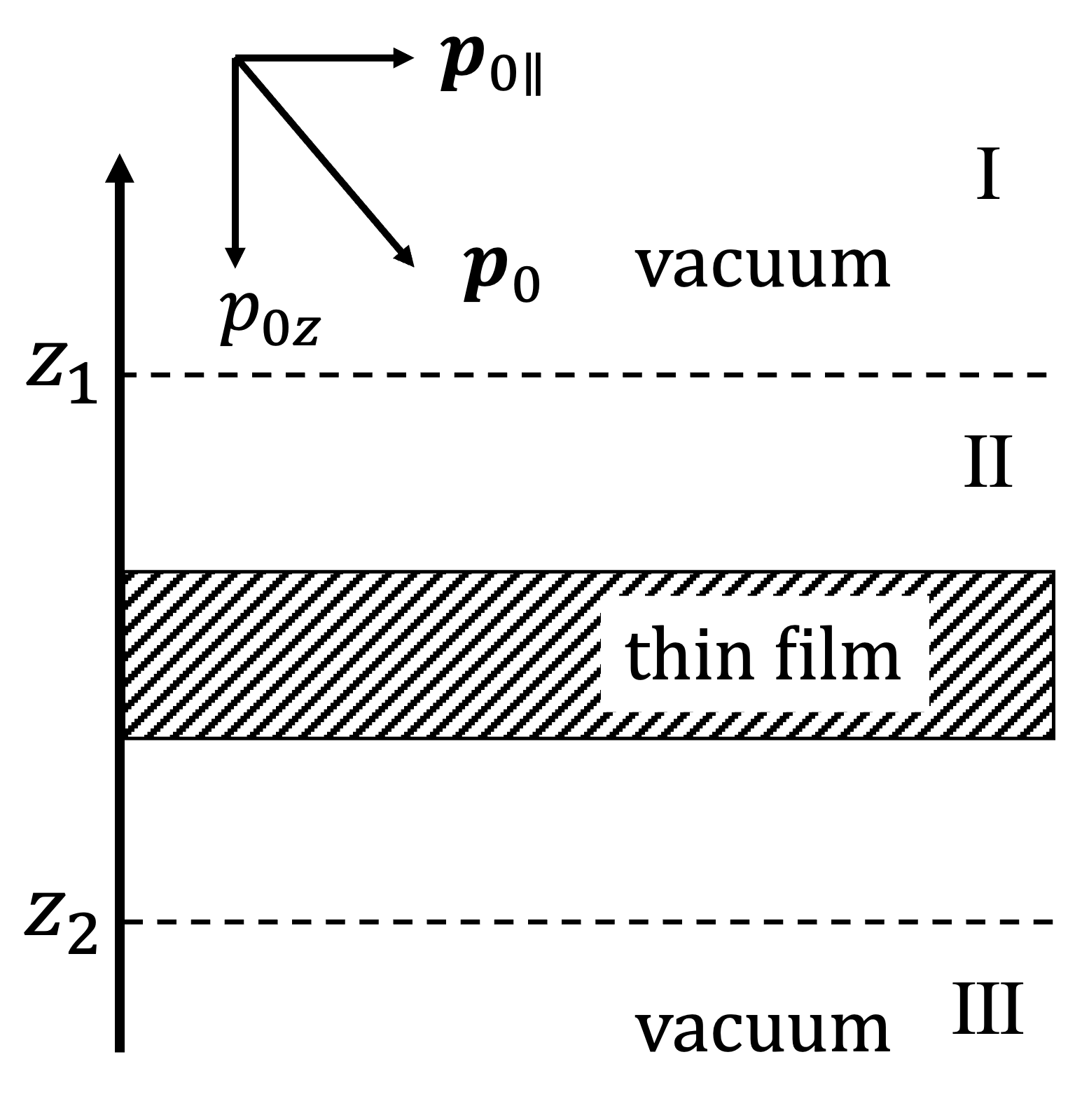}}
\end{tabular}
\caption{Calculation geometry of LEED state $\psi_{{\bm p}_0}$.
}
\label{fig:setup}
\end{figure}

Let us consider free-electron eigenstates with momentum
${\bm p}=({\bm p}_\parallel,p_z)$.
Their energy is $\epsilon_{\bm p}=p^2/2$ and
the surface parallel crystal momentum is
${\bm k}_\parallel={\bm p}_\parallel-{\bm G}_{\parallel}$.
Here ${\bm k}_\parallel$ lies in the first surface Brillouin zone
and ${\bm G}_{\parallel}$ is a surface reciprocal lattice vector.
We have $p_z=\pm q$ where
\begin{eqnarray}
q \equiv q({\bm G}_{\parallel}) \equiv\sqrt{p^2 - \left|{\bm k}_{\parallel} + {\bm G}_{\parallel}\right|^2}
\label{eq:q}
\end{eqnarray}
We divide space in three regions, see Fig.\ref{fig:setup}.
Regions I ($z>z_1$) and III ($z<z_2)$ contain only vacuum,
while region~II ($z_2<z<z_1$) contains the slab and
will be modeled using a supercell of height $c=|z_2-z_1|$.

We consider a LEED state $\psi_{{\bm p}_0}$
where an electron with momentum ${\bm p}_{0}$
is incident from above (region I in Fig.~\ref{fig:setup}).
We have
${\bm p}_{0}=({\bm k}_{\parallel} + {\bm G}_{0\parallel},-q_0)$,
where $q_0\equiv q({\bm G}_{0\parallel})$.
Since the energy and ${\bm k}_{\parallel}$ are conserved,
the wave function in the vacuum regions can be written as
a linear combination of plane waves with momentum
${\bm p}=({\bm k}_{\parallel}+{\bm G}_{\parallel},\pm q)$,
where $+q$ and $-q$ correspond to regions I and III,
respectively and $p=p_0$ is fixed.
Thus we can write the LEED wave as
\begin{equation}
\psi_{\rm I}({\bm r}) = e^{i{\bm k}_{\parallel}\cdot {\bm r}_{\parallel}}
\left(
e^{i {\bm G}_{0\parallel}\cdot {\bm r}_{\parallel}-iq_0z}
  + \sum_{ {\bm G}_{\parallel}} R_{{\bm G}_{\parallel}}
  e^{i {\bm G}_{\parallel}\cdot {\bm r}_{\parallel}+iqz}
  \right) \label{eq:upper}
\end{equation}
\begin{equation}
  \psi_{\rm I\!I\!I}({\bm r}) = e^{i{\bm k}_{\parallel}\cdot {\bm r}_{\parallel}}
\sum_{ {\bm G}_{\parallel}} T_{{\bm G}_{\parallel}}
e^{i {\bm G}_{\parallel}\cdot {\bm r}_{\parallel} -i q z} \label{eq:lower}
\end{equation}
Here the sums over ${\bm G}_{\parallel}$ are restricted by
$|{\bm k}_{\parallel} + {\bm G}_{\parallel}| \le p_0$,
and $R_{{\bm G}_{\parallel}}$,
$T_{{\bm G}_{\parallel}}$ are complex coefficients.
In region~II we develop
the LEED wave over eigenstates 
of the RSG calculation, i.e. Bloch waves
$\varphi_{n{\bm k}}$ with energy
$\epsilon_{n{\bm k}}$,
where ${\bm k}=({\bm k}_{\parallel},k_z)$ and
$n$ is a band index.
The states $\varphi_{n{\bm k}}$ are taken to be
normalized in the supercell volume.
Upon introducing the 3$D$-cell periodic functions,
\begin{eqnarray}
  u_{n{\bm k}}({\bm r}) =\varphi_{n{\bm k}}({\bm r})
  e^{-i{\bm k}\cdot {\bm r}}
  \label{eq:middla}
\end{eqnarray}
we may write
\begin{equation}
  \psi_{\rm I\!I}({\bm r}) = e^{i{\bm k}_{\parallel}\cdot {\bm r}_{\parallel}}
  \sum_{n,k_z}^{\prime} c_{nk_z} u_{n{\bm k}}({\bm r})e^{ik_zz}
\label{eq:middle}
\end{equation}
Here $c_{nk_z}$ are complex coefficients and
the primed sum means that we include only eigenstates
that approximately respect energy conservation,
i.e. $\epsilon_{n{\bm k}} \approx \epsilon_{{\bm p}_0}$.
The coefficients $c_{nk_z}$, $R_{{\bm G}_{\parallel}}$ and $T_{{\bm G}_{\parallel}}$ are determined by the condition that
the wave function and its gradient be continuous at the boundary surfaces,
i.e. at all points (${\bm r}_\parallel,z_1)$ and $({\bm r}_\parallel,z_2)$.
These conditions are more conveniently applied in
reciprocal space. For a 2$D$-cell periodic functions $f({\bm r}_{\parallel})$,
we define the lattice Fourier transform as
\begin{eqnarray}
  f({\bm r}_{\parallel})
 & =& \sum_{ {\bm G}_{\parallel}}{\bar f}_{{\bm G}_{\parallel}}
e^{i{\bm G}_{\parallel}\cdot{\bm r}_{\parallel}}\label{eq:fofr}\\
{\bar f}_{{\bm G}_{\parallel}}
&=&\frac{1}{A}\iint_A d{\bm r}_{\parallel} 
f({\bm r}_{\parallel})e^{-i{\bm G}_{\parallel}\cdot{\bm r}_{\parallel}}
\label{eq:fofg}
\end{eqnarray}
where the integral is over the 2$D$ unit cell with area~$A$.
We have
\begin{eqnarray}
u_{n{\bm k}}({\bm r}_\parallel,z)
= \sum_{ {\bm G}_{\parallel}}
{\bar u}_{n{\bm k};{\bm G}_{\parallel}}(z)
e^{i{\bm G}_{\parallel}\cdot{\bm r}_{\parallel}}
\end{eqnarray}
From Eqs \eqref{eq:upper}\eqref{eq:lower}\eqref{eq:middle} together
with the orthogonality of the functions $e^{i{\bm G}_{\parallel}\cdot{\bm r}_{\parallel}}$, it is
easy to see that the continuity of $\psi$ across the region boundaries
requires that
\begin{eqnarray}
 R_{{\bm G}_{\parallel}} e^{i q z_1}
 &+&\delta_{{\bm G}_{\parallel},{\bm G}_{0\parallel}} e^{-iq_0z_1}\nonumber\\
 &=&\sum_{n,k_z}^{\prime} 
 c_{nk_z} {\bar u}_{n{\bm k};{\bm G}_{\parallel}}(z_1)e^{i k_z z_1} \label{eq:me1}\\
 T_{{\bm G}_{\parallel}} e^{-i q z_2}
&=& \sum_{n,k_z}^{\prime} 
c_{nk_z} {\bar u}_{n{\bm k};{\bm G}_{\parallel}}(z_2)e^{i k_z z_2}\label{eq:me2}
\end{eqnarray}
where $q=q({\bm G}_\parallel)$ is given by Eq.~\eqref{eq:q}.
From the continuity of $\partial\psi/\partial z$ we obtain
\begin{eqnarray}
 R_{{\bm G}_{\parallel}} qe^{i q z_1}
 &-&\delta_{{\bm G}_{\parallel},{\bm G}_{0\parallel}} qe^{-iq_0z_1}\nonumber \\
 &=&\sum_{n,k_z}^{\prime} 
 c_{nk_z} {\bar u}_{n{\bm k};{\bm G}_{\parallel}}(z_1)k_ze^{i k_z z_1} \label{eq:me3}\\
-T_{{\bm G}_{\parallel}} qe^{-i q z_2}
&=& \sum_{n,k_z}^{\prime} 
c_{nk_z} {\bar u}_{n{\bm k};{\bm G}_{\parallel}}(z_2)k_ze^{i k_z z_2}
\label{eq:me4}
\end{eqnarray}
We refer to Eqs~(\ref{eq:me1}--\ref{eq:me4}) as the matching equations.
Note that $R_{{\bm G}_{\parallel}}=T_{{\bm G}_{\parallel}}=0$
for all ${\bm G}_{\parallel}$ with
$|{\bm k}_\parallel+{\bm G}_{\parallel}|>p_0$.
For exact matching, these equations
must hold simultaneously for all ${\bm G}_{\parallel}$,
which leads to an over-determined, infinite dimensional linear problem
for the unknown coefficients, namely the $c_{nk_z}$,
and the $T_{{\bm G}_{\parallel}}$, $R_{{\bm G}_{\parallel}}$
for $|{\bm k}_\parallel+{\bm G}_{\parallel}|<p_0$.
In practice, we retain a finite number of equations for
${\bm G}_{\parallel}$ vectors below some cut-off kinetic energy
and find the least squares solution of the over-determined system.
When a plane wave basis set is used for the RSG calculation, we have
$u_{n{\bm k}}({\bm r})=\sum_{\bm G}C^{n{\bm k}}_{\bm G}e^{i{\bm G}\cdot{\bm r}}$ and
${\bar u}_{n{\bm k};{\bm G}_\parallel}(z)=\sum_{G_z}C^{n{\bm k}}_{\bm G}e^{iG_zz}$ where $C^{n{\bm k}}_{\bm G}$ are the usual plane wave coefficients.
The matrix elements in Eq. \eqref{eq:fermi} are evaluated with 
all-electron wave functions $\varphi_{n{\bm k}}({\bm r})$
or, for comparison,
with pseudo wave functions $\tilde{\varphi}_{n{\bm k}}({\bm r})$,
see the S.M. for details~\cite{supplmat}.
We use the Vienna ab initio simulation package (VASP)~\cite{vasp}
for the RSG calculation and the {\tt VaspUnfolding}
program \cite{VaspUnfolding} for the reconstruction of
the all-electron functions from the PAW pseudo functions~\cite{kressejoubert}.
Note that for the matching problem, Eqs.\eqref{eq:me1}--\eqref{eq:me4},
all-electron functions are not needed, since the matching is done 
at $z=z_{1,2}$, outside any PAW augmentation sphere, where
$\tilde{\varphi}_{n{\bm k}}({\bm r})=\varphi_{n{\bm k}}({\bm r})$.

\section{Application to graphene}
In this section, we apply the method to photoemission from an isolated graphene
sheet, and we study the normal photoemission intensity as function of
photon energy for different initial bands.
First, we calculate the LEED state
$\psi_{{\bm p}_{0}}$ with ${\bm p}_0 = (0,0, p_{0z})$
for $p_{0z}=-\sqrt{2\epsilon_f}$ and then 
the photoemission matrix element Eq. \eqref{eq:fermi} with
the time-reversed LEED state Eq. \eqref{eq:TR}. 

\subsection{Density functional theory calculation}
\begin{figure}[htb]
\centering
\includegraphics[bb= 0 0 504 576, width=0.85\linewidth]{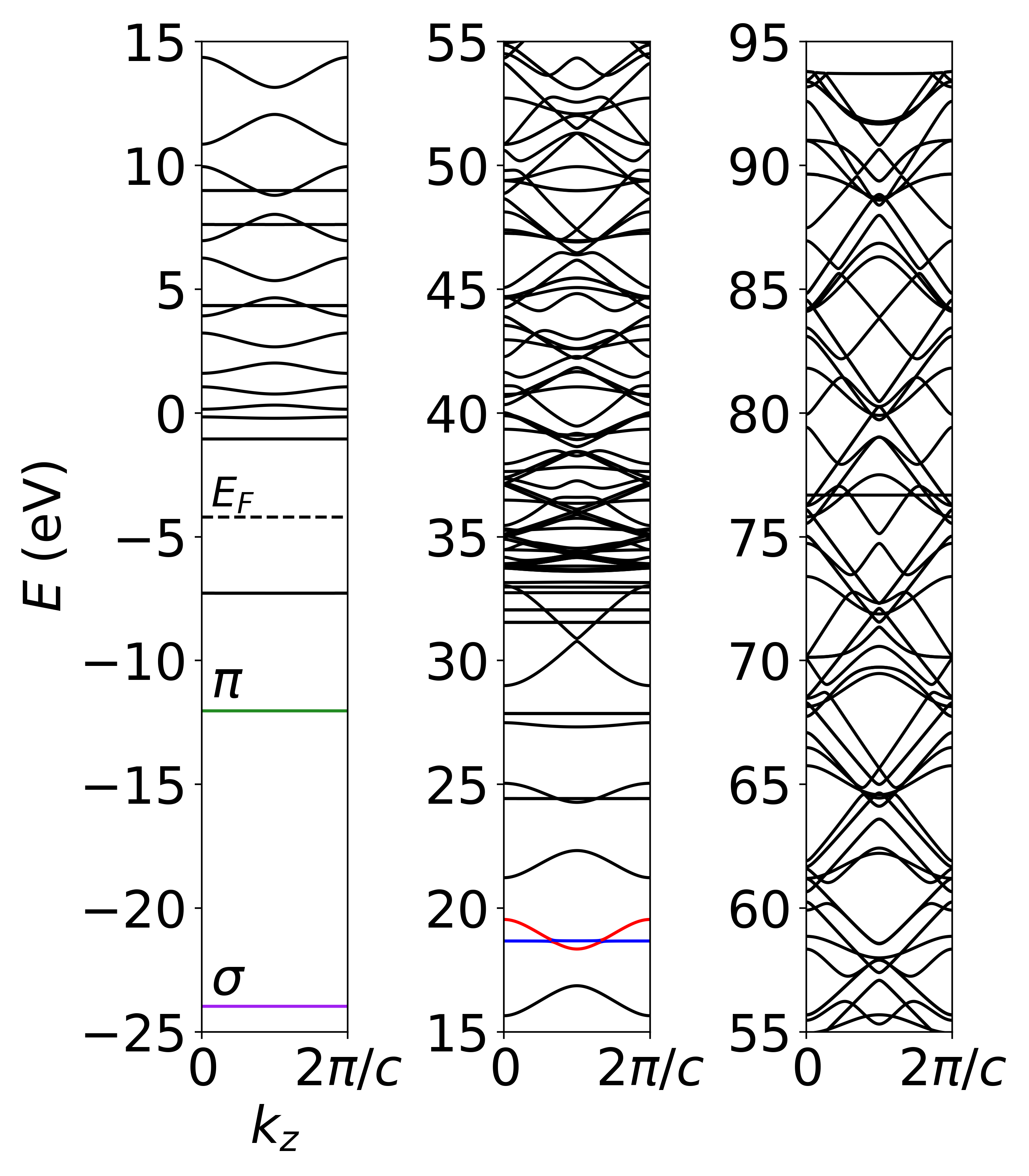}
\caption{$k_z$ band dispersion for ${\bm k}_{\parallel}=(0,0)$.
$E$ is the energy measured from the vacuum level. $c$ is the supercell dimension (20~{\AA}).
The black dashed line indicates Fermi energy ($E_F$).  $\sigma$ and $\pi$ bands are colored purple and green, respectively. Flat and dispersive energy bands around 19 eV are colored blue and red, respectively.
}
\label{fig:band}
\end{figure}

We compute the photoemission
initial states $\psi_i$ and the high energy band states
$\varphi_{n{\bm k}}$ of Eq.~\eqref{eq:middla} in RSG
using the PAW code VASP~\cite{vasp}.
The supercell lattice parameters are $a = 2.44$~{\AA}, $c = 20$~{\AA}.
We use a plane-wave cut-off energy of $400$~eV and a 9$\times$9 $k_\parallel$-mesh 
for the 2$D$ Brillouin zone sampling. Theses are typical settings
for a DFT ground state calculation.
What differs from a usual RSG ground state calculation, is that
we use a fine $k_z$ point mesh with 120 points and
need to compute enough bands up to the wanted photoemission energy.
For the matching equations~\eqref{eq:me1}-\eqref{eq:me4},
400 ${\bm G}_\parallel$ points are used.
%
%
Figure \ref{fig:band} shows the band structure along $k_z$ at ${\bm k}_{\parallel}$=(0,0). Here and in the following $E$ denotes the electron
energy measured from the vacuum level.
The computed work function is 4.21 eV.

\subsection{Construction of LEED states}
In a calculation using RSG, some eigenstates~$\varphi_{n{\bm k}}$ with positive energy
are evanescent and do not propagate along~$z$.
Such states do not contribute to the photoemission process and must be
discarded~\cite{Kob20}.
To identify these states, 
we compute the probability current along~$z$, which is given by
\begin{eqnarray}
J_{z,m} =
\iint_A d{\bm r}_{\parallel} \Im \left [ {\varphi}_{m}^* \frac{\partial}{\partial z} {\varphi}_{m} \right ]
\label{eq:intjz}
\end{eqnarray}
where $m=n{\bm k}$ and $\varphi_{m}({\bm r})$
is evaluated at $z=z_1$.
Details of the flux calculation are given
in section~II of the S.M.~\cite{supplmat}.
We keep only states $\varphi_{n{\bm k}}$
whose flux exceeds some numerical threshold. Here we take
$|J_{z,m}|>3\times10^{-5}$ [a.u.].
For constructing a photoemission final state of energy $\epsilon_f$
one would ideally use only band states with exactly the same energy,
$\epsilon_{n{\bm k}}=\epsilon_f$.
However, for numerical reasons, such as the discretization of $k_z$ space,
we need to slightly relax this constraint and
solve the matching equations with band states
$\varphi_{n{\bm k}_{\parallel}}$ in a small interval around the
exact energy~$\epsilon_{f}$. Here we use
$|\epsilon_{n{\bm k}}-\epsilon_{f}|<0.01$ eV.

Some numerical solutions of the matching equations do not 
correspond to physical scattering states. 
Acceptable LEED states satisfy the probability conservation
\begin{equation}
  {\cal T}+{\cal R} =1,
\label{eq:kakurituhozon}
\end{equation}
where ${\cal T}$ and ${\cal R}$ are the total transmission and reflection coefficients, respectively.
Solutions that do not satisfy Eq.~\eqref{eq:kakurituhozon} are discarded.
For the 2$D$ system we have
\begin{eqnarray}
 {\cal T} &=& {\cal T}_0 + {\cal T}_{\neq 0},\nonumber \\
  {\cal T}_0 &=& |T_{{\bm G}_{0\parallel}}|^2,  \quad 
  {\cal T}_{\neq0}=\sum_{{\bm G}_{\parallel}(\neq {\bm G}_{0\parallel})}\frac{q({\bm G}_{\parallel})}{q_{0}} |T_{{\bm G}_{\parallel}}|^2,
\nonumber \\
 {\cal R} &=& {\cal R}_0 + {\cal R}_{\neq 0},\nonumber \\
{\cal R}_0 &=& |R_{{\bm G}_{0\parallel}}|^2,\quad 
 {\cal R}_0=\sum_{{\bm G}_{\parallel}(\neq {\bm G}_{0\parallel})}\frac{q({\bm G}_{\parallel})}{q_0} |R_{{\bm G}_{\parallel}}|^2.
\end{eqnarray}
with $q({\bm G}_{\parallel})$ defined in Eq.~\eqref{eq:q} and $q_0\equiv q({\bm G}_{0\parallel})$.
To enforce the probability conservation numerically,
we exclude all states with $|{\cal T}+{\cal R}-1| > \eta$,
where $\eta = 0.05$ in the present case. For larger supercell sizes,
smaller $\eta$ values can be used as we have checked.

\begin{figure}[htb]
\centering
\includegraphics[bb= 0 0 640 366, width=0.9\linewidth]{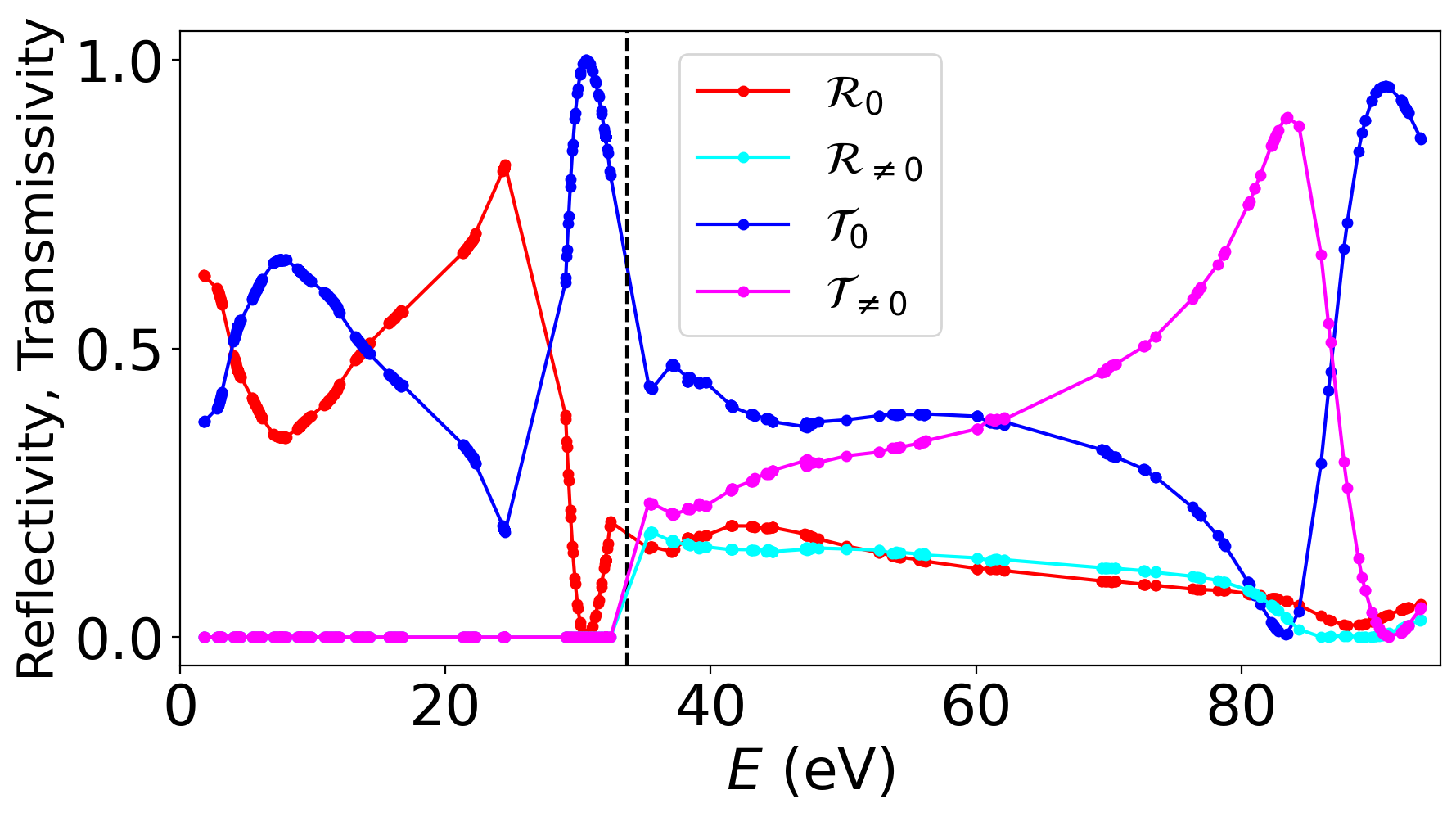}
\caption{
Reflection and transmission coefficients of the calculated LEED states, 
decomposed into ${\bm G}_{\parallel}={\bm G}_{0\parallel}$ and
${\bm G}_{\parallel}\neq{\bm G}_{0\parallel}$ components
(umklapp scattering).
The dashed line indicates the theoretical
onset of umklapp scattering ($E= 33.7$~eV).
The curves are ${\cal R}_0$ (red), ${\cal R}_{\neq 0}$ (cyan),
${\cal T}_0$ (blue) and ${\cal T}_{\neq0}$ (magenta).}
\label{fig:TR}
\end{figure}

The transmission and reflection coefficients are shown
in Fig. \ref{fig:TR}, where the ${\bm G}_{\parallel}=0$
components and the ${\bm G}_{\parallel}\ne 0$
components are plotted separately
in order to illustrate the effect of umklapp scattering (${\bm G}\ne 0$).
Umklapp scattering is possible for $E>{\bm b}_{1}^2/2=8\pi^2/3a^2=33.7$~eV,
indicated by a dotted line in Fig.~\ref{fig:TR}. Above this energy,
direct and umklapp components of the reflectivity are of about equal
intensity, while in transmission, umklapp scattering strongly dominates
for energies around 80~eV.
The energy dependence of the total transmission coefficient
agrees well with Ref.~\cite{nazarov13}.

For some energies, acceptable LEED states cannot be found.
As a result, the plots in Fig. \ref{fig:TR} are not continuous, i.e. there
are small gaps in the energy mesh.
The main reason for this is that the band structure $\epsilon(k_z)$
(Fig.~\ref{fig:band}) has gaps which are due to the artificial
periodicity in the RSG along the $c$ axis.
If the photoelectron energy falls into a gap, the ARPES
result can be obtained by interpolating between two nearby energy points.
If interpolation is deemed not accurate enough,
then one should perform a new RSG calculation with a somewhat
different supercell lattice constant~$c'$.
As it is well known from the Kronig-Penney model~\cite{ashcroft},
a change in lattice constant shifts the energy gaps in the band structure,
and $c'$ can be tuned to obtain a solution for the desired energy.
This is shown in the S.M.~\cite{supplmat} where results for a twice
larger supercell ($c'=40$~\AA) are presented.

\begin{figure}[htb]
\centering
\includegraphics[bb=0 0 344 564, width=0.7\linewidth]{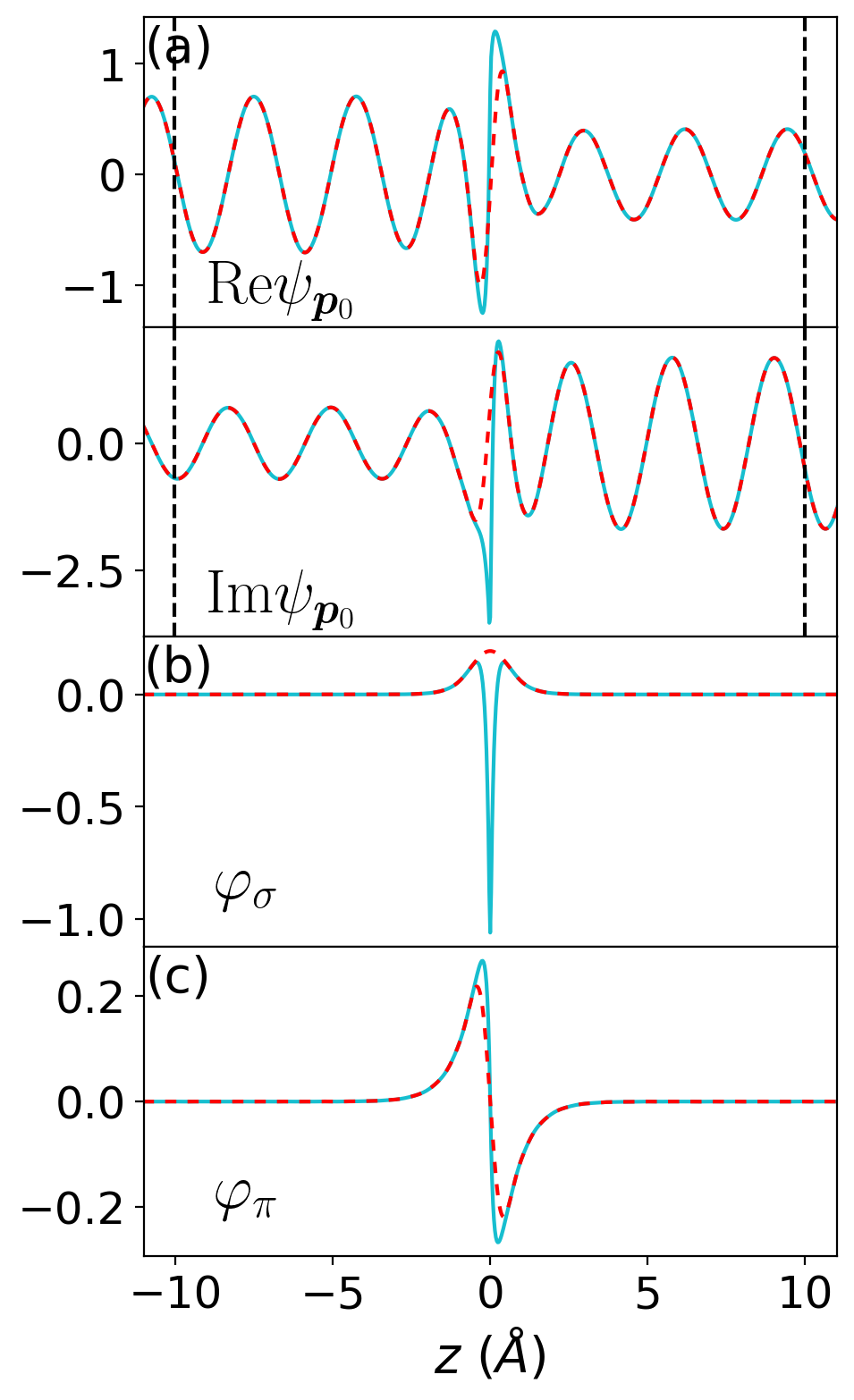}
\caption{Wave function line plots. The lines are along $z$ and cross the center of a C atom at $z=0$.
All-electron waves are shown as cyan solid lines and pseudo waves as red dashed lines.
(a) LEED state at kinetic energy $E = 14$~eV, (b) $\sigma$ initial state
and (c) $\pi$ initial state.
The wave amplitudes are given in units of 1 in (a) and in a.u. in (b,c).
}
\label{fig:leedstate2}
\end{figure}

As an example of the LEED wave functions obtained with the present method,
Fig.~\ref{fig:leedstate2}(a) shows the all-electron wave
$\psi_{{\bm p}_0}$ and the corresponding pseudo wave $\tilde{\psi}_{{\bm p}_0}$ for $\epsilon_f= 14$~eV.
Matching was performed at $z_1 = 10$~{\AA} and $z_2 = -10$~{\AA} (black dashed line).
The obtained wave functions are smooth at these points.
From the behavior of the wave function in Fig.~\ref{fig:leedstate2}(a)
it is clear that the precise choice of the matching points
is not important. Smooth matching can be achieved for wide range of
values, typically $|z_{1,2}|>5$~\AA.

\subsection{Photoemission intensity calculation}
We have calculated the energy dependence of the normal photoemission intensity emitted from graphene with linear $z$-polarized light, i.e.
using Eq. \eqref{eq:fermi} with ${\bm A}={\hat{\bm e}}_z$.
Because of ${\bm k}_{\parallel}$ conservation, we only probe
the $\Gamma$ point in the 2$D$ Brillouin zone.
Among the four valence bands of graphene~\cite{kruger22},
only two have non-zero normal emission intensity for this polarization,
namely the lowest $\sigma$~band of C-2$s$ orbital character
and the $\pi$-band of C-2$p_z$ character, which we refer to
as $\sigma$ and $\pi$ in the following.
Their energy levels are indicated by purple and green lines in Fig. \ref{fig:band}
and wave function plots are displayed in Fig. \ref{fig:leedstate2}(b,c).  

Figure \ref{fig:photoint1}(a) shows the photoemission intensity for the $\sigma$ and
$\pi$ initial states, as a function of photoelectron kinetic energy~$E$.
In both cases, the photoemission intensity has a pronounced energy dependence
including resonances. For the $\sigma$ initial state,
the photoemission intensity grows slowly with energy
from threshold to 85~eV, except for some oscillations between 24~eV and 35~eV.
Since the photoemission final state has time-reversed LEED boundary conditions,
peculiar features of the LEED energy dependence may also
show up in the photoemission spectrum.
We observe that the $\sigma$ photoemission spectral shape
resembles the reflectivity for energies below the on-set of umklapp scattering (33.7~eV)
and the transmissivity for higher energies, see Fig. \ref{fig:TR}.
The peak-dip structures in the LEED reflectivity, around 24--32 eV and 33--35~eV, 
were explained by Nazarov et al.~\cite{nazarov13} as being
caused by resonances in the graphene band structure
and the onset of the umklapp scattering.

The $\pi$ initial state has a very different energy dependence than
the $\sigma$ state. After a rise at threshold, the intensity mostly
decreases from 6~eV to 87~eV. Interestingly there is a very sharp resonance
at $E=19$~eV (i.e. at photon energy $31$~eV)
which has, to the best of our knowledge, not been reported so far.
The resonance is unrelated to the reflectivity (Fig. \ref{fig:TR}).
We have analyzed the final state waves in the RSG calculation at the
2$D$ $\Gamma$-point. The 19~eV resonance corresponds to the lowest
energy ${\bm k}_{\parallel}=0$ state that is both evanescent
and dipole-transition allowed from the $\pi$ initial state.
The energy level of this state is marked in blue in Fig. \ref{fig:band}.
The wave function is evanescent as seen by the absence of $k_z$ dispersion.
It has the same point symmetry as the $\sigma$ initial state
(fully symmetric $a_{1g}$) with a large amplitude inside the C$_6$ rings,
rather than on the C-atoms, see the S.M.~\cite{supplmat}.
A plane-wave-like state propagating in $+z$-direction of the same
energy (the red band in Fig. \ref{fig:band}) has a small overlap with the evanescent state.
This leads a wave function mixing and, as we have checked, to
an avoided crossing of the two bands. (The gap opening is so small
that it cannot be seen on the scale of Fig. \ref{fig:band}.)
The weak coupling between the evanescent state and the
free-electron-like continuum states produces the
sharp Fano-like resonance in the $\pi$-band photoemission,
seen in Fig.~\ref{fig:photoint1}(a) at $E=19$~eV.

In Fig. \ref{fig:photoint1}(b) we compare the ARPES intensity of the $\sigma$ state
obtained with the present (``matching'') method
with two independent theoretical methods, namely
real-space multiple scattering (MS)~\cite{kruger11,kruger22}
and the final state plane-wave (FSPW) approximation.
We note that the real-space MS has been shown to be a reliable 
method for ARPES calculations of various system,
including oriented molecules~\cite{kruger18}, metal
surfaces~\cite{kruger11} and graphite~\cite{kruger22}.
The energy dependence of ``matching'' and ``MS'' is very similar, which
proves the validity of the present matching approach.
The MS calculation shows small differences,
namely a generally more smooth energy dependence and the absence of
a peak at 33.7~eV (dashed vertical line). Both differences can be
attributed to approximations used in the real-space MS method~\cite{kruger11},
namely the finite-cluster approximation which broadens the band features,
and the simple model used for the surface barrier,
which fails to reproduce the ``purely structural'' LEED resonance at the
on-set of umklapp scattering~\cite{nazarov13}.
Apart from these details, the overall agreement between ``matching''	and ``MS''
is very good and it is even better for the $\pi$ initial state [Fig. \ref{fig:photoint1}(c)].
Note that in Fig. \ref{fig:photoint1}(b) and \ref{fig:photoint1}(c),
we used the same intensity scaling for ``matching'' and ``MS''.
It follows that not only the spectral shapes but also the
relative intensity between $\sigma$ and $\pi$
agrees between ``matching'' and ``MS''.
The FSPW on the other hand,
gives very poor results, see blue curves in Fig. \ref{fig:photoint1}(b)
and \ref{fig:photoint1}(c). 
For the $\sigma$-band, the energy dependence is totally wrong.
For the $\pi$ band, the overall dependence is better, but the
resonance at 19~eV and the minimum at 85~eV are missing.
We conclude that the FSPW approximation is unreliable for
the photon energy dependence of ARPES, confirming other recent
literature~\cite{Ono18,horseshoe,dauth}.


\subsection{Comparison of pseudo and all-electron wave function}
\begin{figure}[htbp] 
\begin{tabular}{l}
 (a)\\
 \includegraphics[bb=0 0 441 258, width=0.9\linewidth ]{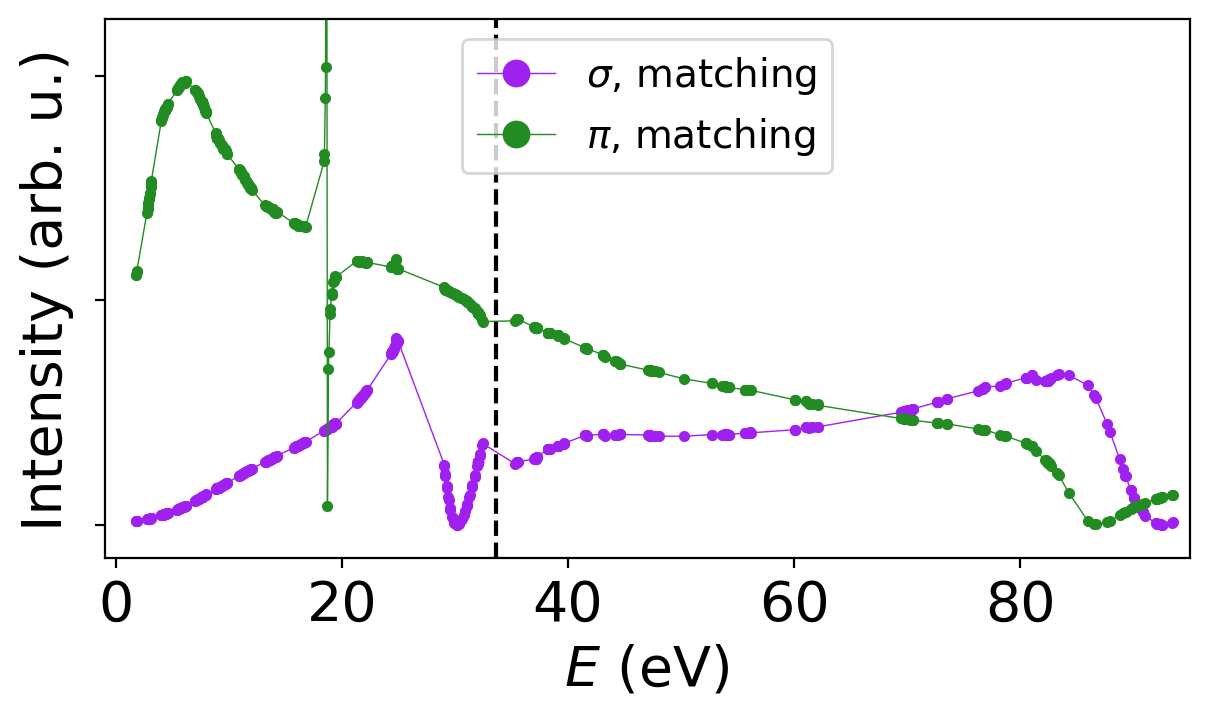} \\
 (b)\\
   \includegraphics[bb=0 0 441 258, width=0.9\linewidth ]{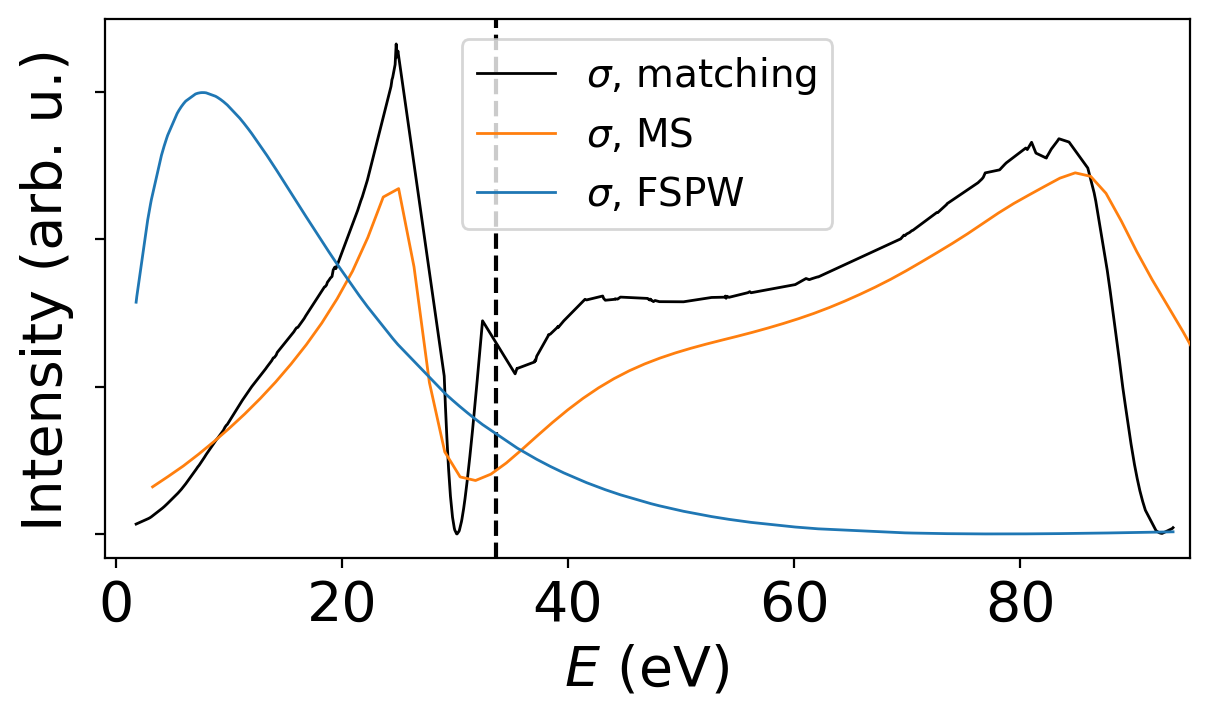}\\
 (c)\\
   \includegraphics[bb=0 0 441 258, width=0.9\linewidth ]{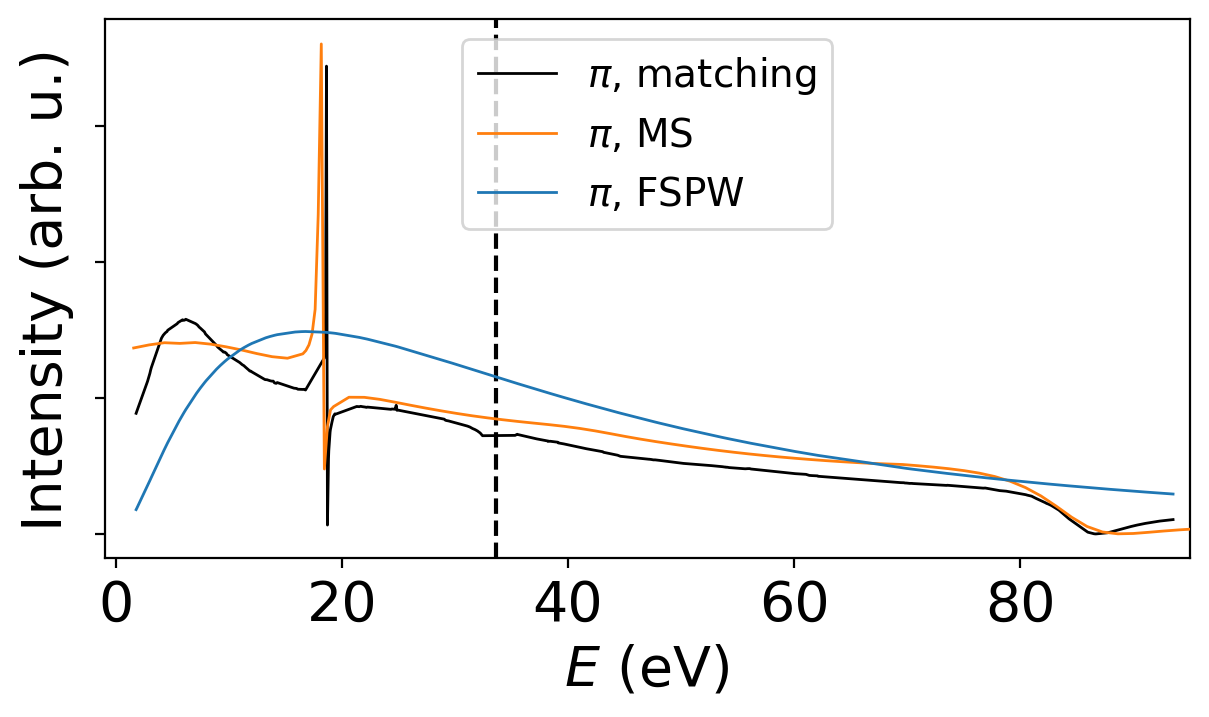}
  \end{tabular}
\caption{
  (a) Normal emission intensity from the graphene $\sigma$ and $\pi$ bands
  obtained with the present ``matching'' method as a function of kinetic energy~$E$.
  (b,c) Comparison with a multiple scattering calculation (MS),
  and the final state plane wave approximation (FSPW).
  In (b) and (c), the calculated intensities are arbitrarily scaled keeping the ratio
  of $\sigma$ and $\pi$ intensities within each method.}
\label{fig:photoint1}
\end{figure}

In the calculation of Fig. \ref{fig:photoint1}(a), we used all-electron wave functions for both initial and final states in the photoemission intensity calculations. Here, we discuss the question
whether we all-electron wave functions can be replaced with pseudo wave functions.

\begin{figure}[htbp] 
  \begin{tabular}{l}
   (a)\\
   \includegraphics[bb=0 0 441 258, width=0.9\linewidth ]{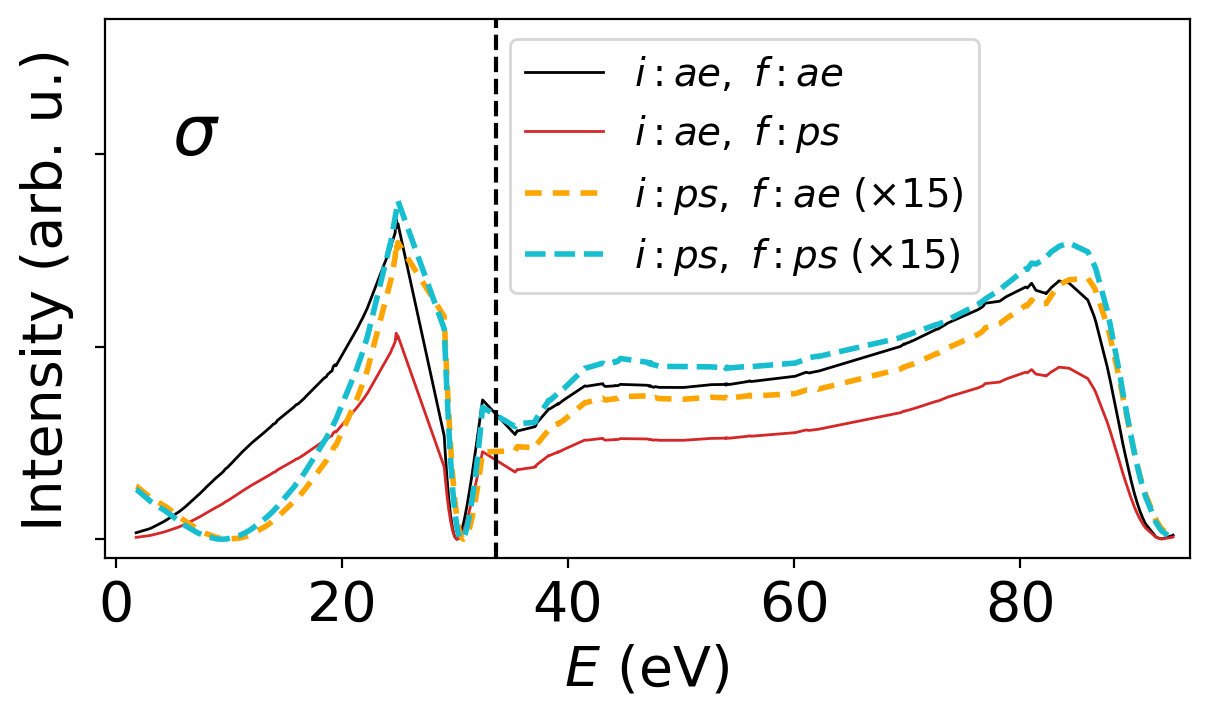} \\
   (b)\\
     \includegraphics[bb=0 0 441 258, width=0.9\linewidth ]{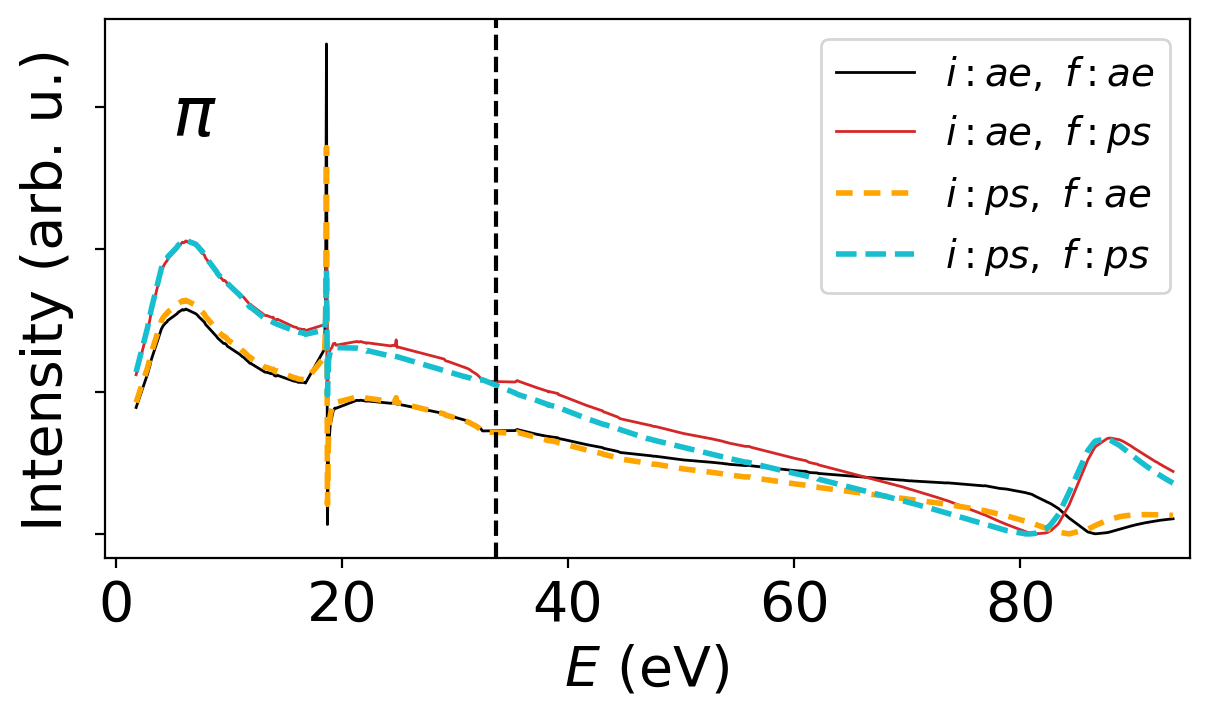}\\
  \end{tabular}
  \caption{Comparison between all-electron (ae) and pseudo (ps) wave functions
    for initial (i) and final (f) states, in the matching method
    calculation of normal photoemission from graphene.
    (a) $\sigma$ initial state. (b) $\pi$ initial state.
    Note that the intensity of the (i:ps) spectra in (a)
    are multiplied by a factor~15.
}  \label{fig:photoint2}
  \end{figure}

In Figure \ref{fig:photoint2}(a,b), we compare four types of calculation 
by choosing either the all-electron (``ae'') or the pseudo (``ps'') wave function for the initial and the
final state.
The overall energy dependence is similar between the four calculation schemes but there are
also important differences.
This is particularly true for the $\sigma$ initial state, Fig.~\ref{fig:photoint2}(a).
When replacing the all-electron wave with a pseudo-wave in the final state,
the photoemission intensity may change by up to 30\%.
When the pseudo wave is used in for the initial state, then the intensity is reduced by
one order of magnitude as compared to the all-electron calculation (note the
scaling factor $\times 15$ used in the plot).
Moreover, the energy dependence is very different for $E<20$~eV.

The differences are smaller for the $\pi$ initial state, Fig.~\ref{fig:photoint2}(b).
Replacing the all-electron initial state wave by the pseudo wave hardly changes the intensity
for $E<60$~eV, although some changes are seen for higher energy.
If the all-electron functions are replaced by pseudo functions in
the final state,
the photoemission intensity is reduced by about 20\% for $E<80$~eV and strongly increased for $E>85$~eV.
The reason why the choice between all-electron and pseudo function has
a much larger effect for the $\sigma$
band than for the $\pi$ band, is obvious when looking at the wave function plots in
Fig.~\ref{fig:leedstate2}(b,c).
The $\sigma$ pseudo wave has no node, while the all-electron
wave has the node of the C-2$s$ atomic wave.

Looking at the similar spectral shapes of Fig\ref{fig:photoint2}(a,b)
one might think that using pseudo wave functions can be an acceptable
approximation when total intensities are not relevant.
However, this is misleading because here we have
considered only the $\Gamma$-point,
where C-$2s$ and C-$2p$ orbitals cannot mix by symmetry.
For a general ${\bm k}_{\parallel}$-point, the $\sigma$ band contains C-$2s$ and C-$2p_{xy}$ components.
Since the $2s$ pseudo-wave leads to a dramatic decrease of intensity but the $2p$ pseudo-wave does
not, there is no way to rescale the pseudo-wave intensity to match the correct all-electron intensity.
We conclude that, in general, all-electron wave functions must be used for a
reliable photoemission intensity calculation. A possible exception
might be the $\pi$-states of graphene and
flat organic molecules~\cite{puschnig}, as
these states are of purely C-$p$ character.

\section{Conclusions}
In summary, we have developed a flexible and accurate computational method for ARPES from 2$D$
systems. The photoemission final states are obtained by
matching Bloch waves of a DFT calculation in repeated-slab-geometry
with free waves having the proper time-reversed LEED boundary conditions. 
The DFT calculation is performed in the PAW method with a standard code.
The method has been applied to ARPES from a graphene monolayer.
The results were compared with independently performed multiple scattering
calculations~\cite{kruger22} and very good agreement was obtained in all cases.
The closely related LEED reflectivity spectrum was also checked against the theoretical
literature~\cite{Kra04}.
In the PAW method, transition matrix elements can be calculated either with PAW pseudo-waves or
with reconstructed all-electron waves. The comparison shows that all-electron
waves must be used for obtaining reliable ARPES intensities.
We have studied the energy dependence of the normal emission intensity from the graphene
$\sigma$ and $\pi$ bands
in the kinetic energy range 0--95~eV. For both bands, we find a pronounced energy dependence with
resonances. We predict a sharp Fano resonance for $\pi$-band normal emission
at a kinetic energy around 19~eV, due to a strong 2$D$ interband transition which is
weakly coupled to a free-electron-like final state.
\bibliography{ref} 

\begin{thebibliography}{27}%
\makeatletter
\providecommand \@ifxundefined [1]{%
 \@ifx{#1\undefined}
}%
\providecommand \@ifnum [1]{%
 \ifnum #1\expandafter \@firstoftwo
 \else \expandafter \@secondoftwo
 \fi
}%
\providecommand \@ifx [1]{%
 \ifx #1\expandafter \@firstoftwo
 \else \expandafter \@secondoftwo
 \fi
}%
\providecommand \natexlab [1]{#1}%
\providecommand \enquote  [1]{``#1''}%
\providecommand \bibnamefont  [1]{#1}%
\providecommand \bibfnamefont [1]{#1}%
\providecommand \citenamefont [1]{#1}%
\providecommand \href@noop [0]{\@secondoftwo}%
\providecommand \href [0]{\begingroup \@sanitize@url \@href}%
\providecommand \@href[1]{\@@startlink{#1}\@@href}%
\providecommand \@@href[1]{\endgroup#1\@@endlink}%
\providecommand \@sanitize@url [0]{\catcode `\\12\catcode `\$12\catcode
  `\&12\catcode `\#12\catcode `\^12\catcode `\_12\catcode `\%12\relax}%
\providecommand \@@startlink[1]{}%
\providecommand \@@endlink[0]{}%
\providecommand \url  [0]{\begingroup\@sanitize@url \@url }%
\providecommand \@url [1]{\endgroup\@href {#1}{\urlprefix }}%
\providecommand \urlprefix  [0]{URL }%
\providecommand \Eprint [0]{\href }%
\providecommand \doibase [0]{https://doi.org/}%
\providecommand \selectlanguage [0]{\@gobble}%
\providecommand \bibinfo  [0]{\@secondoftwo}%
\providecommand \bibfield  [0]{\@secondoftwo}%
\providecommand \translation [1]{[#1]}%
\providecommand \BibitemOpen [0]{}%
\providecommand \bibitemStop [0]{}%
\providecommand \bibitemNoStop [0]{.\EOS\space}%
\providecommand \EOS [0]{\spacefactor3000\relax}%
\providecommand \BibitemShut  [1]{\csname bibitem#1\endcsname}%
\let\auto@bib@innerbib\@empty
\bibitem [{\citenamefont {Damascelli}\ \emph {et~al.}(2003)\citenamefont
  {Damascelli}, \citenamefont {Hussain},\ and\ \citenamefont {Shen}}]{ARPES}%
  \BibitemOpen
  \bibfield  {author} {\bibinfo {author} {\bibfnamefont {A.}~\bibnamefont
  {Damascelli}}, \bibinfo {author} {\bibfnamefont {Z.}~\bibnamefont
  {Hussain}},\ and\ \bibinfo {author} {\bibfnamefont {Z.-X.}\ \bibnamefont
  {Shen}},\ }\bibfield  {title} {\bibinfo {title} {{Angle-resolved
  photoemission studies of the cuprate superconductors}},\ }\href
  {https://doi.org/10.1103/RevModPhys.75.473} {\bibfield  {journal} {\bibinfo
  {journal} {Rev. Mod. Phys.}\ }\textbf {\bibinfo {volume} {75}},\ \bibinfo
  {pages} {473} (\bibinfo {year} {2003})}\BibitemShut {NoStop}%
\bibitem [{\citenamefont {Puschnig}\ \emph {et~al.}(2009)\citenamefont
  {Puschnig}, \citenamefont {Berkebile}, \citenamefont {Fleming}, \citenamefont
  {Koller}, \citenamefont {Emtsev}, \citenamefont {Seyller}, \citenamefont
  {Riley}, \citenamefont {Ambrosch-Draxl}, \citenamefont {Netzer},\ and\
  \citenamefont {Ramsey}}]{puschnig}%
  \BibitemOpen
  \bibfield  {author} {\bibinfo {author} {\bibfnamefont {P.}~\bibnamefont
  {Puschnig}}, \bibinfo {author} {\bibfnamefont {S.}~\bibnamefont {Berkebile}},
  \bibinfo {author} {\bibfnamefont {A.~J.}\ \bibnamefont {Fleming}}, \bibinfo
  {author} {\bibfnamefont {G.}~\bibnamefont {Koller}}, \bibinfo {author}
  {\bibfnamefont {K.}~\bibnamefont {Emtsev}}, \bibinfo {author} {\bibfnamefont
  {T.}~\bibnamefont {Seyller}}, \bibinfo {author} {\bibfnamefont {J.~D.}\
  \bibnamefont {Riley}}, \bibinfo {author} {\bibfnamefont {C.}~\bibnamefont
  {Ambrosch-Draxl}}, \bibinfo {author} {\bibfnamefont {F.~P.}\ \bibnamefont
  {Netzer}},\ and\ \bibinfo {author} {\bibfnamefont {M.~G.}\ \bibnamefont
  {Ramsey}},\ }\bibfield  {title} {\bibinfo {title} {{Reconstruction of
  Molecular Orbital Densities from Photoemission Data}},\ }\href
  {https://doi.org/10.1126/science.1176105} {\bibfield  {journal} {\bibinfo
  {journal} {Science}\ }\textbf {\bibinfo {volume} {326}},\ \bibinfo {pages}
  {702} (\bibinfo {year} {2009})}\BibitemShut {NoStop}%
\bibitem [{\citenamefont {Tusche}\ \emph {et~al.}(2019)\citenamefont {Tusche},
  \citenamefont {Chen}, \citenamefont {Schneider},\ and\ \citenamefont
  {Kirschner}}]{PMM}%
  \BibitemOpen
  \bibfield  {author} {\bibinfo {author} {\bibfnamefont {C.}~\bibnamefont
  {Tusche}}, \bibinfo {author} {\bibfnamefont {Y.-J.}\ \bibnamefont {Chen}},
  \bibinfo {author} {\bibfnamefont {C.~M.}\ \bibnamefont {Schneider}},\ and\
  \bibinfo {author} {\bibfnamefont {J.}~\bibnamefont {Kirschner}},\ }\bibfield
  {title} {\bibinfo {title} {{Imaging properties of hemispherical electrostatic
  energy analyzers for high resolution momentum microscopy}},\ }\href
  {https://doi.org/https://doi.org/10.1016/j.ultramic.2019.112815} {\bibfield
  {journal} {\bibinfo  {journal} {Ultramicroscopy}\ }\textbf {\bibinfo {volume}
  {206}},\ \bibinfo {pages} {112815} (\bibinfo {year} {2019})}\BibitemShut
  {NoStop}%
\bibitem [{\citenamefont {Pendry}(1976)}]{pendry}%
  \BibitemOpen
  \bibfield  {author} {\bibinfo {author} {\bibfnamefont {J.}~\bibnamefont
  {Pendry}},\ }\bibfield  {title} {\bibinfo {title} {{Theory of
  photoemission}},\ }\href
  {https://doi.org/https://doi.org/10.1016/0039-6028(76)90355-1} {\bibfield
  {journal} {\bibinfo  {journal} {Surface Science}\ }\textbf {\bibinfo {volume}
  {57}},\ \bibinfo {pages} {679} (\bibinfo {year} {1976})}\BibitemShut
  {NoStop}%
\bibitem [{\citenamefont {Braun}(1996)}]{braun}%
  \BibitemOpen
  \bibfield  {author} {\bibinfo {author} {\bibfnamefont {J.}~\bibnamefont
  {Braun}},\ }\bibfield  {title} {\bibinfo {title} {{The theory of
  angle-resolved ultraviolet photoemission and its applications to ordered
  materials}},\ }\href {https://doi.org/10.1088/0034-4885/59/10/002} {\bibfield
   {journal} {\bibinfo  {journal} {Reports on Progress in Physics}\ }\textbf
  {\bibinfo {volume} {59}},\ \bibinfo {pages} {1267} (\bibinfo {year}
  {1996})}\BibitemShut {NoStop}%
\bibitem [{\citenamefont {Ono}\ \emph {et~al.}(2021)\citenamefont {Ono},
  \citenamefont {Marmodoro}, \citenamefont {Schusser}, \citenamefont {Nakata},
  \citenamefont {Schwier}, \citenamefont {Braun}, \citenamefont {Ebert},
  \citenamefont {Min\'ar}, \citenamefont {Sakamoto},\ and\ \citenamefont
  {Kr\"uger}}]{ono21}%
  \BibitemOpen
  \bibfield  {author} {\bibinfo {author} {\bibfnamefont {R.}~\bibnamefont
  {Ono}}, \bibinfo {author} {\bibfnamefont {A.}~\bibnamefont {Marmodoro}},
  \bibinfo {author} {\bibfnamefont {J.}~\bibnamefont {Schusser}}, \bibinfo
  {author} {\bibfnamefont {Y.}~\bibnamefont {Nakata}}, \bibinfo {author}
  {\bibfnamefont {E.~F.}\ \bibnamefont {Schwier}}, \bibinfo {author}
  {\bibfnamefont {J.}~\bibnamefont {Braun}}, \bibinfo {author} {\bibfnamefont
  {H.}~\bibnamefont {Ebert}}, \bibinfo {author} {\bibfnamefont
  {J.}~\bibnamefont {Min\'ar}}, \bibinfo {author} {\bibfnamefont
  {K.}~\bibnamefont {Sakamoto}},\ and\ \bibinfo {author} {\bibfnamefont
  {P.}~\bibnamefont {Kr\"uger}},\ }\bibfield  {title} {\bibinfo {title}
  {Surface band characters of the weyl semimetal candidate material
  ${\mathrm{mote}}_{2}$ revealed by one-step angle-resolved photoemission
  theory},\ }\href {https://doi.org/10.1103/PhysRevB.103.125139} {\bibfield
  {journal} {\bibinfo  {journal} {Phys. Rev. B}\ }\textbf {\bibinfo {volume}
  {103}},\ \bibinfo {pages} {125139} (\bibinfo {year} {2021})}\BibitemShut
  {NoStop}%
\bibitem [{\citenamefont {Antonios~Gonis}(2000)}]{gonisbutler}%
  \BibitemOpen
  \bibfield  {author} {\bibinfo {author} {\bibfnamefont {W.~H.~B.}\
  \bibnamefont {Antonios~Gonis}},\ }\href
  {https://doi.org/https://doi.org/10.1007/978-1-4612-1290-4} {\emph {\bibinfo
  {title} {Multiple Scattering in Solids}}},\ \bibinfo {edition} {1st}\ ed.,\
  Graduate Texts in Contemporary Physics\ (\bibinfo  {publisher}
  {Springer-Verlag, New York},\ \bibinfo {year} {2000})\BibitemShut {NoStop}%
\bibitem [{\citenamefont {Hatada}\ \emph {et~al.}(2007)\citenamefont {Hatada},
  \citenamefont {Hayakawa}, \citenamefont {Benfatto},\ and\ \citenamefont
  {Natoli}}]{keisuke}%
  \BibitemOpen
  \bibfield  {author} {\bibinfo {author} {\bibfnamefont {K.}~\bibnamefont
  {Hatada}}, \bibinfo {author} {\bibfnamefont {K.}~\bibnamefont {Hayakawa}},
  \bibinfo {author} {\bibfnamefont {M.}~\bibnamefont {Benfatto}},\ and\
  \bibinfo {author} {\bibfnamefont {C.~R.}\ \bibnamefont {Natoli}},\ }\bibfield
   {title} {\bibinfo {title} {{Full-potential multiple scattering for x-ray
  spectroscopies}},\ }\href {https://doi.org/10.1103/PhysRevB.76.060102}
  {\bibfield  {journal} {\bibinfo  {journal} {Phys. Rev. B}\ }\textbf {\bibinfo
  {volume} {76}},\ \bibinfo {pages} {060102} (\bibinfo {year}
  {2007})}\BibitemShut {NoStop}%
\bibitem [{\citenamefont {Bl\"ochl}(1994)}]{blochlpaw}%
  \BibitemOpen
  \bibfield  {author} {\bibinfo {author} {\bibfnamefont {P.~E.}\ \bibnamefont
  {Bl\"ochl}},\ }\bibfield  {title} {\bibinfo {title} {{Projector
  augmented-wave method}},\ }\href {https://doi.org/10.1103/PhysRevB.50.17953}
  {\bibfield  {journal} {\bibinfo  {journal} {Phys. Rev. B}\ }\textbf {\bibinfo
  {volume} {50}},\ \bibinfo {pages} {17953} (\bibinfo {year}
  {1994})}\BibitemShut {NoStop}%
\bibitem [{\citenamefont {Krasovskii}(2004)}]{Kra04}%
  \BibitemOpen
  \bibfield  {author} {\bibinfo {author} {\bibfnamefont {E.~E.}\ \bibnamefont
  {Krasovskii}},\ }\bibfield  {title} {\bibinfo {title} {{Augmented-plane-wave
  approach to scattering of Bloch electrons by an interface}},\ }\href
  {https://doi.org/10.1103/PhysRevB.70.245322} {\bibfield  {journal} {\bibinfo
  {journal} {Phys. Rev. B}\ }\textbf {\bibinfo {volume} {70}},\ \bibinfo
  {pages} {245322} (\bibinfo {year} {2004})}\BibitemShut {NoStop}%
\bibitem [{\citenamefont {Krasovskii}(2021)}]{Kra21}%
  \BibitemOpen
  \bibfield  {author} {\bibinfo {author} {\bibfnamefont {E.}~\bibnamefont
  {Krasovskii}},\ }\bibfield  {title} {\bibinfo {title} {{Ab Initio Theory of
  Photoemission from Graphene}},\ }\bibfield  {journal} {\bibinfo  {journal}
  {Nanomaterials}\ }\textbf {\bibinfo {volume} {11}},\ \href
  {https://doi.org/10.3390/nano11051212} {10.3390/nano11051212} (\bibinfo
  {year} {2021})\BibitemShut {NoStop}%
\bibitem [{\citenamefont {Krasovskii}(2020)}]{Kra20}%
  \BibitemOpen
  \bibfield  {author} {\bibinfo {author} {\bibfnamefont {E.~E.}\ \bibnamefont
  {Krasovskii}},\ }\bibfield  {title} {\bibinfo {title} {Character of the
  outgoing wave in soft x-ray photoemission},\ }\href
  {https://doi.org/10.1103/PhysRevB.102.245139} {\bibfield  {journal} {\bibinfo
   {journal} {Phys. Rev. B}\ }\textbf {\bibinfo {volume} {102}},\ \bibinfo
  {pages} {245139} (\bibinfo {year} {2020})}\BibitemShut {NoStop}%
\bibitem [{\citenamefont {Kobayashi}(2020)}]{Kob20}%
  \BibitemOpen
  \bibfield  {author} {\bibinfo {author} {\bibfnamefont {K.}~\bibnamefont
  {Kobayashi}},\ }\bibfield  {title} {\bibinfo {title} {{Method of forming
  time-reversed LEED states from repeated-slab calculations}},\ }\href
  {https://doi.org/10.1088/1361-648X/abb444} {\bibfield  {journal} {\bibinfo
  {journal} {Journal of Physics: Condensed Matter}\ }\textbf {\bibinfo {volume}
  {32}},\ \bibinfo {pages} {495002} (\bibinfo {year} {2020})}\BibitemShut
  {NoStop}%
\bibitem [{\citenamefont {Kresse}\ and\ \citenamefont
  {Joubert}(1999)}]{kressejoubert}%
  \BibitemOpen
  \bibfield  {author} {\bibinfo {author} {\bibfnamefont {G.}~\bibnamefont
  {Kresse}}\ and\ \bibinfo {author} {\bibfnamefont {D.}~\bibnamefont
  {Joubert}},\ }\bibfield  {title} {\bibinfo {title} {{From ultrasoft
  pseudopotentials to the projector augmented-wave method}},\ }\href
  {https://doi.org/10.1103/PhysRevB.59.1758} {\bibfield  {journal} {\bibinfo
  {journal} {Phys. Rev. B}\ }\textbf {\bibinfo {volume} {59}},\ \bibinfo
  {pages} {1758} (\bibinfo {year} {1999})}\BibitemShut {NoStop}%
\bibitem [{\citenamefont {Ono}\ and\ \citenamefont {Kr\"uger}(2018)}]{Ono18}%
  \BibitemOpen
  \bibfield  {author} {\bibinfo {author} {\bibfnamefont {R.}~\bibnamefont
  {Ono}}\ and\ \bibinfo {author} {\bibfnamefont {P.}~\bibnamefont {Kr\"uger}},\
  }\bibfield  {title} {\bibinfo {title} {{A One-Dimensional Model for
  Photoemission Calculations from Plane-Wave Band Structure Codes}},\ }\href
  {https://doi.org/10.1380/ejssnt.2018.49} {\bibfield  {journal} {\bibinfo
  {journal} {e-Journal of Surface Science and Nanotechnology}\ }\textbf
  {\bibinfo {volume} {16}},\ \bibinfo {pages} {49} (\bibinfo {year}
  {2018})}\BibitemShut {NoStop}%
\bibitem [{\citenamefont {Dauth}\ \emph {et~al.}(2016)\citenamefont {Dauth},
  \citenamefont {Graus}, \citenamefont {Schelter}, \citenamefont {Wie\ss{}ner},
  \citenamefont {Sch\"oll}, \citenamefont {Reinert},\ and\ \citenamefont
  {K\"ummel}}]{dauth}%
  \BibitemOpen
  \bibfield  {author} {\bibinfo {author} {\bibfnamefont {M.}~\bibnamefont
  {Dauth}}, \bibinfo {author} {\bibfnamefont {M.}~\bibnamefont {Graus}},
  \bibinfo {author} {\bibfnamefont {I.}~\bibnamefont {Schelter}}, \bibinfo
  {author} {\bibfnamefont {M.}~\bibnamefont {Wie\ss{}ner}}, \bibinfo {author}
  {\bibfnamefont {A.}~\bibnamefont {Sch\"oll}}, \bibinfo {author}
  {\bibfnamefont {F.}~\bibnamefont {Reinert}},\ and\ \bibinfo {author}
  {\bibfnamefont {S.}~\bibnamefont {K\"ummel}},\ }\bibfield  {title} {\bibinfo
  {title} {{Perpendicular Emission, Dichroism, and Energy Dependence in
  Angle-Resolved Photoemission: The Importance of The Final State}},\ }\href
  {https://doi.org/10.1103/PhysRevLett.117.183001} {\bibfield  {journal}
  {\bibinfo  {journal} {Phys. Rev. Lett.}\ }\textbf {\bibinfo {volume} {117}},\
  \bibinfo {pages} {183001} (\bibinfo {year} {2016})}\BibitemShut {NoStop}%
\bibitem [{\citenamefont {Kern}\ \emph {et~al.}(2023)\citenamefont {Kern},
  \citenamefont {Haags}, \citenamefont {Egger}, \citenamefont {Yang},
  \citenamefont {Kirschner}, \citenamefont {Wolff}, \citenamefont {Seyller},
  \citenamefont {Gottwald}, \citenamefont {Richter}, \citenamefont
  {De~Giovannini}, \citenamefont {Rubio}, \citenamefont {Ramsey}, \citenamefont
  {Bocquet}, \citenamefont {Soubatch}, \citenamefont {Tautz}, \citenamefont
  {Puschnig},\ and\ \citenamefont {Moser}}]{horseshoe}%
  \BibitemOpen
  \bibfield  {author} {\bibinfo {author} {\bibfnamefont {C.~S.}\ \bibnamefont
  {Kern}}, \bibinfo {author} {\bibfnamefont {A.}~\bibnamefont {Haags}},
  \bibinfo {author} {\bibfnamefont {L.}~\bibnamefont {Egger}}, \bibinfo
  {author} {\bibfnamefont {X.}~\bibnamefont {Yang}}, \bibinfo {author}
  {\bibfnamefont {H.}~\bibnamefont {Kirschner}}, \bibinfo {author}
  {\bibfnamefont {S.}~\bibnamefont {Wolff}}, \bibinfo {author} {\bibfnamefont
  {T.}~\bibnamefont {Seyller}}, \bibinfo {author} {\bibfnamefont
  {A.}~\bibnamefont {Gottwald}}, \bibinfo {author} {\bibfnamefont
  {M.}~\bibnamefont {Richter}}, \bibinfo {author} {\bibfnamefont
  {U.}~\bibnamefont {De~Giovannini}}, \bibinfo {author} {\bibfnamefont
  {A.}~\bibnamefont {Rubio}}, \bibinfo {author} {\bibfnamefont {M.~G.}\
  \bibnamefont {Ramsey}}, \bibinfo {author} {\bibfnamefont {F.~m. c.~C.}\
  \bibnamefont {Bocquet}}, \bibinfo {author} {\bibfnamefont {S.}~\bibnamefont
  {Soubatch}}, \bibinfo {author} {\bibfnamefont {F.~S.}\ \bibnamefont {Tautz}},
  \bibinfo {author} {\bibfnamefont {P.}~\bibnamefont {Puschnig}},\ and\
  \bibinfo {author} {\bibfnamefont {S.}~\bibnamefont {Moser}},\ }\bibfield
  {title} {\bibinfo {title} {{Simple extension of the plane-wave final state in
  photoemission: Bringing understanding to the photon-energy dependence of
  two-dimensional materials}},\ }\href
  {https://doi.org/10.1103/PhysRevResearch.5.033075} {\bibfield  {journal}
  {\bibinfo  {journal} {Phys. Rev. Res.}\ }\textbf {\bibinfo {volume} {5}},\
  \bibinfo {pages} {033075} (\bibinfo {year} {2023})}\BibitemShut {NoStop}%
\bibitem [{\citenamefont {Kasmi}\ \emph {et~al.}(2017)\citenamefont {Kasmi},
  \citenamefont {Lucchini}, \citenamefont {Castiglioni}, \citenamefont
  {Kliuiev}, \citenamefont {Osterwalder}, \citenamefont {Hengsberger},
  \citenamefont {Gallmann}, \citenamefont {Kr\"{u}ger},\ and\ \citenamefont
  {Keller}}]{optica}%
  \BibitemOpen
  \bibfield  {author} {\bibinfo {author} {\bibfnamefont {L.}~\bibnamefont
  {Kasmi}}, \bibinfo {author} {\bibfnamefont {M.}~\bibnamefont {Lucchini}},
  \bibinfo {author} {\bibfnamefont {L.}~\bibnamefont {Castiglioni}}, \bibinfo
  {author} {\bibfnamefont {P.}~\bibnamefont {Kliuiev}}, \bibinfo {author}
  {\bibfnamefont {J.}~\bibnamefont {Osterwalder}}, \bibinfo {author}
  {\bibfnamefont {M.}~\bibnamefont {Hengsberger}}, \bibinfo {author}
  {\bibfnamefont {L.}~\bibnamefont {Gallmann}}, \bibinfo {author}
  {\bibfnamefont {P.}~\bibnamefont {Kr\"{u}ger}},\ and\ \bibinfo {author}
  {\bibfnamefont {U.}~\bibnamefont {Keller}},\ }\bibfield  {title} {\bibinfo
  {title} {Effective mass effect in attosecond electron transport},\ }\href
  {https://doi.org/10.1364/OPTICA.4.001492} {\bibfield  {journal} {\bibinfo
  {journal} {Optica}\ }\textbf {\bibinfo {volume} {4}},\ \bibinfo {pages}
  {1492} (\bibinfo {year} {2017})}\BibitemShut {NoStop}%
\bibitem [{\citenamefont {Brown}\ \emph {et~al.}(1980)\citenamefont {Brown},
  \citenamefont {Carter},\ and\ \citenamefont {Kelly}}]{Brown80}%
  \BibitemOpen
  \bibfield  {author} {\bibinfo {author} {\bibfnamefont {E.~R.}\ \bibnamefont
  {Brown}}, \bibinfo {author} {\bibfnamefont {S.~L.}\ \bibnamefont {Carter}},\
  and\ \bibinfo {author} {\bibfnamefont {H.~P.}\ \bibnamefont {Kelly}},\
  }\bibfield  {title} {\bibinfo {title} {{Photoionization cross section and
  resonance structure of ClI}},\ }\href
  {https://doi.org/10.1103/PhysRevA.21.1237} {\bibfield  {journal} {\bibinfo
  {journal} {Phys. Rev. A}\ }\textbf {\bibinfo {volume} {21}},\ \bibinfo
  {pages} {1237} (\bibinfo {year} {1980})}\BibitemShut {NoStop}%
\bibitem [{sup()}]{supplmat}%
  \BibitemOpen
  \href@noop {} {\ }\bibinfo {note} {See Supplemental Material at \dots for
  details about (I) the calculation of the photoemission matrix elements, (II)
  the probability current $J_{z,m}$, (III) results obtained with a larger
  supercell size $c=40$~\AA\ and (IV) the wave function character at the
  resonance energy $E=19$~eV.}\BibitemShut {Stop}%
\bibitem [{\citenamefont {Kresse}\ and\ \citenamefont
  {Furthm\"uller}(1996)}]{vasp}%
  \BibitemOpen
  \bibfield  {author} {\bibinfo {author} {\bibfnamefont {G.}~\bibnamefont
  {Kresse}}\ and\ \bibinfo {author} {\bibfnamefont {J.}~\bibnamefont
  {Furthm\"uller}},\ }\bibfield  {title} {\bibinfo {title} {{Efficient
  iterative schemes for ab initio total-energy calculations using a plane-wave
  basis set}},\ }\href {https://doi.org/10.1103/PhysRevB.54.11169} {\bibfield
  {journal} {\bibinfo  {journal} {Phys. Rev. B}\ }\textbf {\bibinfo {volume}
  {54}},\ \bibinfo {pages} {11169} (\bibinfo {year} {1996})}\BibitemShut
  {NoStop}%
\bibitem [{Vas()}]{VaspUnfolding}%
  \BibitemOpen
  \href@noop {} {}\bibinfo {note}
  {\url{https://github.com/QijingZheng/VaspBandUnfolding}}\BibitemShut
  {NoStop}%
\bibitem [{\citenamefont {Nazarov}\ \emph {et~al.}(2013)\citenamefont
  {Nazarov}, \citenamefont {Krasovskii},\ and\ \citenamefont
  {Silkin}}]{nazarov13}%
  \BibitemOpen
  \bibfield  {author} {\bibinfo {author} {\bibfnamefont {V.~U.}\ \bibnamefont
  {Nazarov}}, \bibinfo {author} {\bibfnamefont {E.~E.}\ \bibnamefont
  {Krasovskii}},\ and\ \bibinfo {author} {\bibfnamefont {V.~M.}\ \bibnamefont
  {Silkin}},\ }\bibfield  {title} {\bibinfo {title} {Scattering resonances in
  two-dimensional crystals with application to graphene},\ }\href
  {https://doi.org/10.1103/PhysRevB.87.041405} {\bibfield  {journal} {\bibinfo
  {journal} {Phys. Rev. B}\ }\textbf {\bibinfo {volume} {87}},\ \bibinfo
  {pages} {041405} (\bibinfo {year} {2013})}\BibitemShut {NoStop}%
\bibitem [{\citenamefont {Ashcroft}\ and\ \citenamefont
  {Mermin}(1976)}]{ashcroft}%
  \BibitemOpen
  \bibfield  {author} {\bibinfo {author} {\bibfnamefont {N.~W.}\ \bibnamefont
  {Ashcroft}}\ and\ \bibinfo {author} {\bibfnamefont {N.~D.}\ \bibnamefont
  {Mermin}},\ }\href@noop {} {\emph {\bibinfo {title} {Solid State Physics}}}\
  (\bibinfo  {publisher} {Saunders College, Philadelphia},\ \bibinfo {year}
  {1976})\BibitemShut {NoStop}%
\bibitem [{\citenamefont {Krüger}\ and\ \citenamefont
  {Matsui}(2022)}]{kruger22}%
  \BibitemOpen
  \bibfield  {author} {\bibinfo {author} {\bibfnamefont {P.}~\bibnamefont
  {Krüger}}\ and\ \bibinfo {author} {\bibfnamefont {F.}~\bibnamefont
  {Matsui}},\ }\bibfield  {title} {\bibinfo {title} {{Observation and theory of
  strong circular dichroism in angle-revolved photoemission from graphite}},\
  }\href {https://doi.org/https://doi.org/10.1016/j.elspec.2022.147219}
  {\bibfield  {journal} {\bibinfo  {journal} {Journal of Electron Spectroscopy
  and Related Phenomena}\ }\textbf {\bibinfo {volume} {258}},\ \bibinfo {pages}
  {147219} (\bibinfo {year} {2022})}\BibitemShut {NoStop}%
\bibitem [{\citenamefont {Kr\"uger}\ \emph {et~al.}(2011)\citenamefont
  {Kr\"uger}, \citenamefont {Da~Pieve},\ and\ \citenamefont
  {Osterwalder}}]{kruger11}%
  \BibitemOpen
  \bibfield  {author} {\bibinfo {author} {\bibfnamefont {P.}~\bibnamefont
  {Kr\"uger}}, \bibinfo {author} {\bibfnamefont {F.}~\bibnamefont {Da~Pieve}},\
  and\ \bibinfo {author} {\bibfnamefont {J.}~\bibnamefont {Osterwalder}},\
  }\bibfield  {title} {\bibinfo {title} {{Real-space multiple scattering method
  for angle-resolved photoemission and valence-band photoelectron diffraction
  and its application to Cu(111)}},\ }\href
  {https://doi.org/10.1103/PhysRevB.83.115437} {\bibfield  {journal} {\bibinfo
  {journal} {Phys. Rev. B}\ }\textbf {\bibinfo {volume} {83}},\ \bibinfo
  {pages} {115437} (\bibinfo {year} {2011})}\BibitemShut {NoStop}%
\bibitem [{\citenamefont {Kr\"{u}ger}(2018)}]{kruger18}%
  \BibitemOpen
  \bibfield  {author} {\bibinfo {author} {\bibfnamefont {P.}~\bibnamefont
  {Kr\"{u}ger}},\ }\bibfield  {title} {\bibinfo {title} {{Photoelectron
  Diffraction from Valence States of Oriented Molecules}},\ }\href
  {https://doi.org/10.7566/JPSJ.87.061007} {\bibfield  {journal} {\bibinfo
  {journal} {Journal of the Physical Society of Japan}\ }\textbf {\bibinfo
  {volume} {87}},\ \bibinfo {pages} {061007} (\bibinfo {year}
  {2018})}\BibitemShut {NoStop}%
\end{thebibliography}%


\begin{thebibliography}{1}%
\makeatletter
\providecommand \@ifxundefined [1]{%
 \@ifx{#1\undefined}
}%
\providecommand \@ifnum [1]{%
 \ifnum #1\expandafter \@firstoftwo
 \else \expandafter \@secondoftwo
 \fi
}%
\providecommand \@ifx [1]{%
 \ifx #1\expandafter \@firstoftwo
 \else \expandafter \@secondoftwo
 \fi
}%
\providecommand \natexlab [1]{#1}%
\providecommand \enquote  [1]{``#1''}%
\providecommand \bibnamefont  [1]{#1}%
\providecommand \bibfnamefont [1]{#1}%
\providecommand \citenamefont [1]{#1}%
\providecommand \href@noop [0]{\@secondoftwo}%
\providecommand \href [0]{\begingroup \@sanitize@url \@href}%
\providecommand \@href[1]{\@@startlink{#1}\@@href}%
\providecommand \@@href[1]{\endgroup#1\@@endlink}%
\providecommand \@sanitize@url [0]{\catcode `\\12\catcode `\$12\catcode
  `\&12\catcode `\#12\catcode `\^12\catcode `\_12\catcode `\%12\relax}%
\providecommand \@@startlink[1]{}%
\providecommand \@@endlink[0]{}%
\providecommand \url  [0]{\begingroup\@sanitize@url \@url }%
\providecommand \@url [1]{\endgroup\@href {#1}{\urlprefix }}%
\providecommand \urlprefix  [0]{URL }%
\providecommand \Eprint [0]{\href }%
\providecommand \doibase [0]{https://doi.org/}%
\providecommand \selectlanguage [0]{\@gobble}%
\providecommand \bibinfo  [0]{\@secondoftwo}%
\providecommand \bibfield  [0]{\@secondoftwo}%
\providecommand \translation [1]{[#1]}%
\providecommand \BibitemOpen [0]{}%
\providecommand \bibitemStop [0]{}%
\providecommand \bibitemNoStop [0]{.\EOS\space}%
\providecommand \EOS [0]{\spacefactor3000\relax}%
\providecommand \BibitemShut  [1]{\csname bibitem#1\endcsname}%
\let\auto@bib@innerbib\@empty
\bibitem [{\citenamefont {Momma}\ and\ \citenamefont {Izumi}(2011)}]{VESTA}%
  \BibitemOpen
  \bibfield  {author} {\bibinfo {author} {\bibfnamefont {K.}~\bibnamefont
  {Momma}}\ and\ \bibinfo {author} {\bibfnamefont {F.}~\bibnamefont {Izumi}},\
  }\bibfield  {title} {\bibinfo {title} {{VESTA 3 for three-dimensional
  visualization of crystal, volumetric and morphology data}},\ }\href@noop {}
  {\bibfield  {journal} {\bibinfo  {journal} {J. Appl. Crystallogr.}\ }\textbf
  {\bibinfo {volume} {44}},\ \bibinfo {pages} {1272} (\bibinfo {year}
  {2011})}\BibitemShut {NoStop}%
\end{thebibliography}%

\acknowledgements
This work was supported by the Japan Science and Technology Agency (JST),
through a university fellowship for science-technology-innovation, Grant Number JPMJFS2107.
\end{document}


\title{
Supplemental Material for 
``A computational method for angle-resolved photoemission
  spectra from repeated-slab band structure calculations''
}
\author{Misa Nozaki}
\email[]{6d697361@gmail.com}
\author{Peter Kr\"uger}
\email{pkruger@faculty.chiba-u.jp}
\affiliation{Graduate School of Science and Engineering, Chiba University, Yayoi-cho 1-33, Inage, Chiba 263-8522, Japan}
\date{\today}

\begin{abstract}
Details are given about (I) the calculation of the
photoemission matrix elements, (II) the probability current $J_{z,m}$,
(III) results obtained with a larger supercell
size $c=40$~\AA\ and (IV) the wave function character at the
resonance energy $E=19$~eV.
\end{abstract}

\maketitle

\section{Calculation of photoemission matrix elements}
Details of the photoemission matrix element calculation of Eq.~(1) are given
for the case when both initial and final states are expanded in plane waves,
e.g. when pseudo functions ${\tilde\varphi}$ are used.
In the case of all-electron waves $\varphi$,
the integrals in Eq.~\eqref{eq:ps3} are performed numerically.
\begin{eqnarray}
&&\braket{ {\tilde \psi}_{-{\bm p}_0}|{\bm \varepsilon} \cdot {\hat {\bm P}} | {\tilde \varphi}_i } \nonumber \\
&=& -i \sum_{a=x,y,z} \varepsilon_a  
\int_{z_a}^{z_b} dz \iint_A d{\bm r}_{\parallel} {\tilde \psi}_{-{\bm p}_0}^{*}({\bm r}) \frac{\partial}{\partial x_a} {\tilde \varphi}_i({\bm r}) \nonumber \\
&=&-i \sum_{a=x,y,z} \varepsilon_a  \sum_{m} c_m
\int_{z_a}^{z_b} dz \iint_A d{\bm r}_{\parallel} {\tilde \varphi_m}({\bm r}) \frac{\partial}{\partial x_a} {\tilde \varphi}_i({\bm r})
\label{eq:matrix}
\end{eqnarray}
where $m=n{\bf k}$ is a band state, $A$ is the area of the 2$D$ unit cell and
\begin{eqnarray}
{\tilde \psi}_{-{\bm p}_0}^{*} ({\bm r})&=& 
{\tilde \psi}_{{\bm p}_0}({\bm r})
=\sum_{m} c_m {\tilde\varphi}_{m}({\bm r})
,\label{eq:ps1}\\
{\tilde \varphi_m}({\bm r}) &=& \frac{1}{ \sqrt {V_{\rm cell}}}\sum_{{\bm G}} C_{{\bm G}}^m e^{i({\bm G}+{\bm k}_{m})\cdot {\bm r}}.
\label{eq:ps2}
\end{eqnarray}
$z_a$ and $z_b$ are determined as $\varphi_{i}({\bm r})=0$ in $z<z_a$ and $z > z_b$.
Substituting Eq. \eqref{eq:ps1} and Eq. \eqref{eq:ps2} into Eq. \eqref{eq:matrix}, the integral in the last line of Eq. \eqref{eq:matrix} becomes
\begin{eqnarray}
&&\int_{z_a}^{z_b} dz \iint_A d{\bm r}_{\parallel}  
{\tilde \varphi}_{m}(\bm r)\frac{\partial}{\partial a} {\tilde \varphi}_{i}(\bm r)\nonumber \\
&=&\frac{1}{V_{\rm cell}}\sum_{\bm G} \sum_{{\bm G}'} C^i_{\bf G} C_{{\bf G}'}^{m} 
\int_{z_a}^{z_b} dz \iint_A d {\bm r}_{\parallel} i(G_a + k_{ia}) e^{i ( {\bm G} + {\bm G}' + {\bm k}_i + {\bm k}_m ) \cdot {\bm r} } \nonumber \\
&=& c\delta_{{\bf k}_{i\parallel}, -{\bf k}_{m\parallel} } \sum_{ {\bf G}_{\parallel}} \sum_{G_z}\sum_{G_z'} 
C^i_{ {\bf G}_{\parallel}, G_z } C_{-{\bf G}_{\parallel}, G_z'}^{m}  i(G_a  + k_{ia} )
 \int_{z_a}^{z_b} dz  e^{i ( G_z + G_z' + k_{iz} + k_{mz} ) \cdot z } \nonumber\\
&=& 
c\delta_{{\bm k}_{i\parallel}, -{\bm k}_{m\parallel} } \sum_{ {\bm G}_{\parallel}} \sum_{G_z} \sum_{G_z'} 
C^i_{ {\bf G}_{\parallel}, G_z } C_{-{\bf G}_{\parallel}, G_z'}^{m}  i(G_a + k_{ia} ) \nonumber \\
&\times& \left [ \left(1-\delta_{G_z, -G_z'} \delta_{ k_{iz},-k_{mz} } \right)
\frac{e^{i ( G_z + G_z' + k_{iz} + k_{mz} ) \cdot z_2 } - e^{i ( G_z + G_z' + k_{iz} + k_{mz} ) \cdot z_1 } }{i ( G_z + G_z' + k_{iz} + k_{mz} )  }
+ \delta_{G_z, -G_z'} \delta_{ k_{iz},-k_{mz} } (z_b-z_a) \right ].
\label{eq:ps3}
\end{eqnarray}

\clearpage
\section{Calculation of probability current}
\begin{figure}[htb]
  \centering
  \begin{tabular}{c}
    \includegraphics[bb= 0 0 504 576, width=0.4\linewidth]{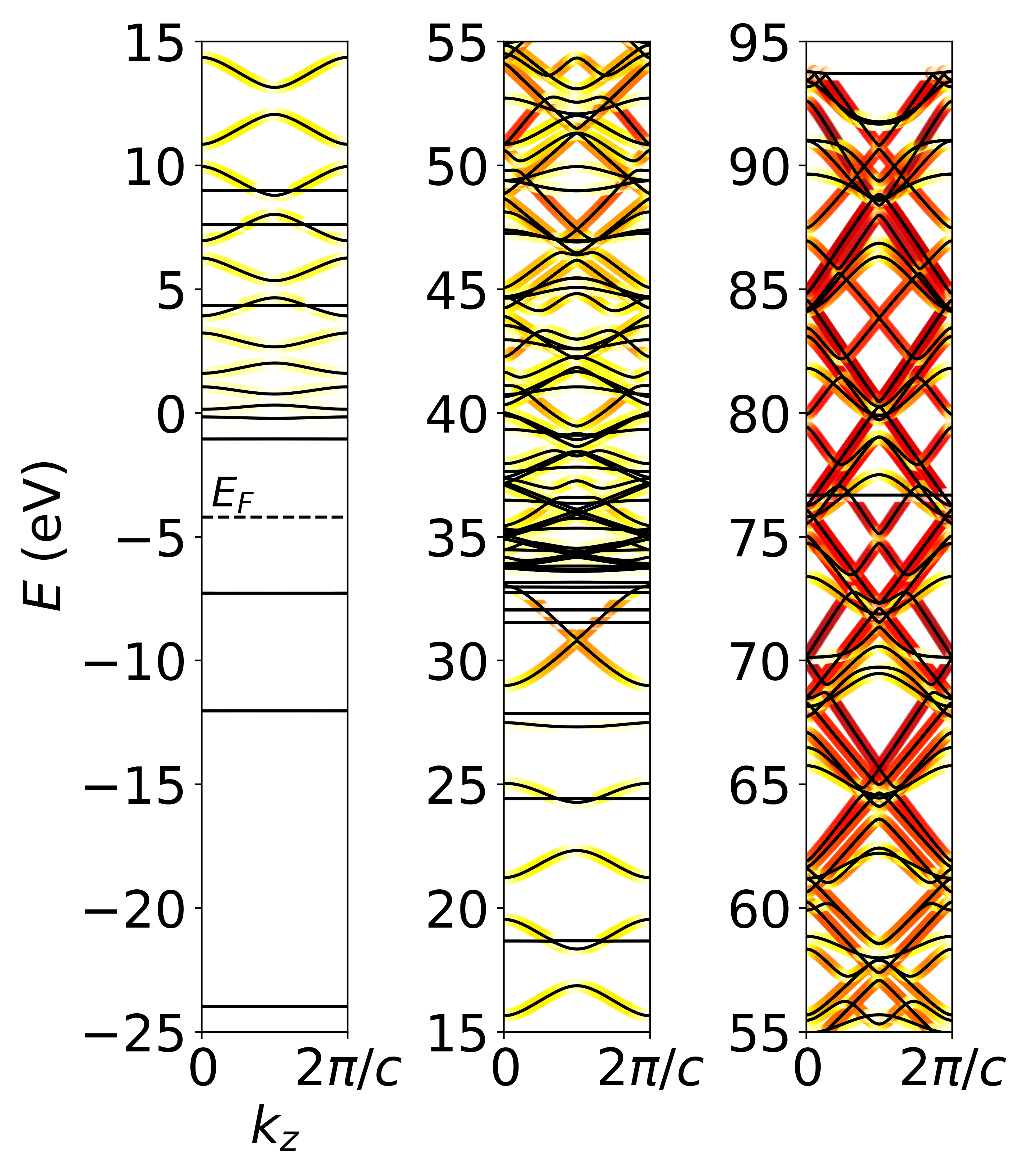}\\
    \includegraphics[bb= 0 0 588 109, width=0.4\linewidth]{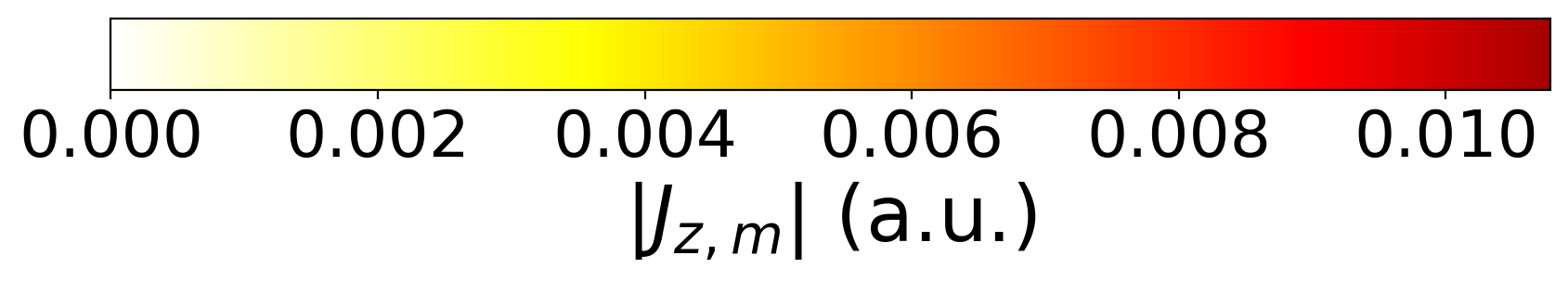}
  \end{tabular}
  \caption{Probability current projected band structure corresponding to Fig. 2.
    $|J_{z,m}|$ is evaluated at $z = 10$ {\AA}. }
  \label{figure:color}
    \end{figure}

The probablitiy current along $z$, given in Eq.~(15) of the main text,
is calculated, using Eq. \eqref{eq:ps2}, as 
\begin{eqnarray}
J_{z, m}(z)
&=&
\iint_A d{\bm r}_{\parallel}  {\rm Im} \left [ {\tilde  \varphi}^*_{m}({\bm r}_{\parallel},z) \frac{\partial  {\tilde \varphi}_{m}({\bm r}_{\parallel},z) }{\partial z} \right ]\nonumber\\
&=&  \frac{1}{V_{\rm cell}}{\rm Im}
\left[
  \sum_{\bm G}\sum_{{\bm G}'} C_{\bm G}^* C_{{\bm G}'} i(k_z+G_z') 
  e^{i(G_z'-G_z)z}
\int \int  d{\bm r}_{\parallel}  e^{i({\bm G}_{\parallel}'-{\bm G}_{\parallel})\cdot{\bm r}_{\parallel}} 
\right ] \nonumber \\
&=&
c\;{\rm Im} 
\left[
 \sum_{{\bm G}_{\parallel}} 
 \sum_{G_z} \sum_{{G_z}'}
 C_{{\bm G}_\parallel, G_z}^* C_{{\bm G}_{\parallel}, G_z'} i(k_z+G_z') e^{i(G_z'-G_z)z}
\right ].
\end{eqnarray}
Fig. \ref{figure:color} shows $|J_{z,m}|$ evaluated at $z$ = 10 {\AA} for all
band states ${\varphi}_{n{\bm k}}$ with ${\bf k}_\parallel=(0,0)$.

\clearpage
\section{Comparison between different supercell sizes}
\begin{figure}[htb]
 \centering
\begin{tabular}{lll}
  (a)&&(b)\\
  \includegraphics[bb= 0 0 425 360, width=0.3\linewidth]{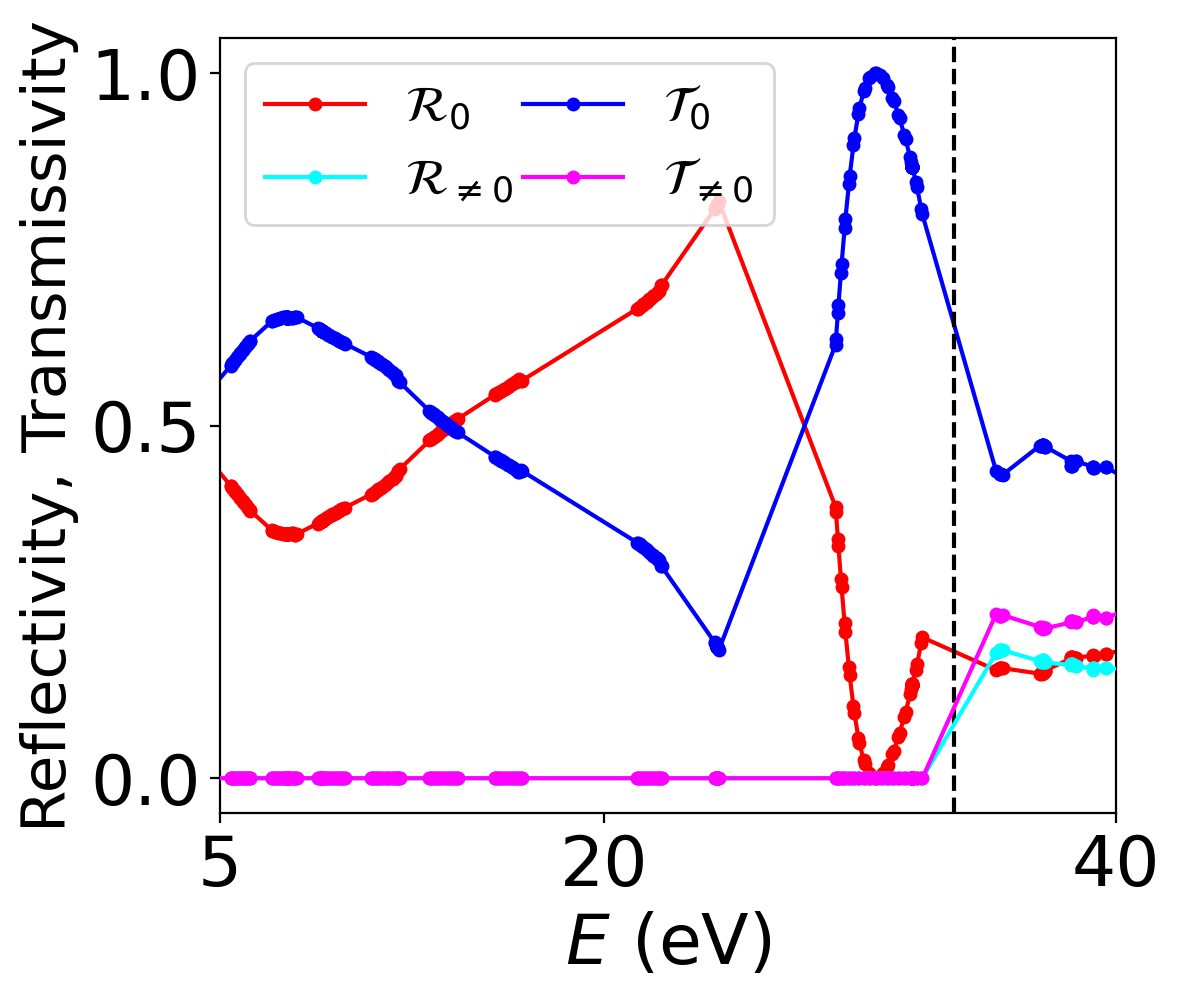}
  && \includegraphics[bb= 0 0 425 360, width=0.3\linewidth]{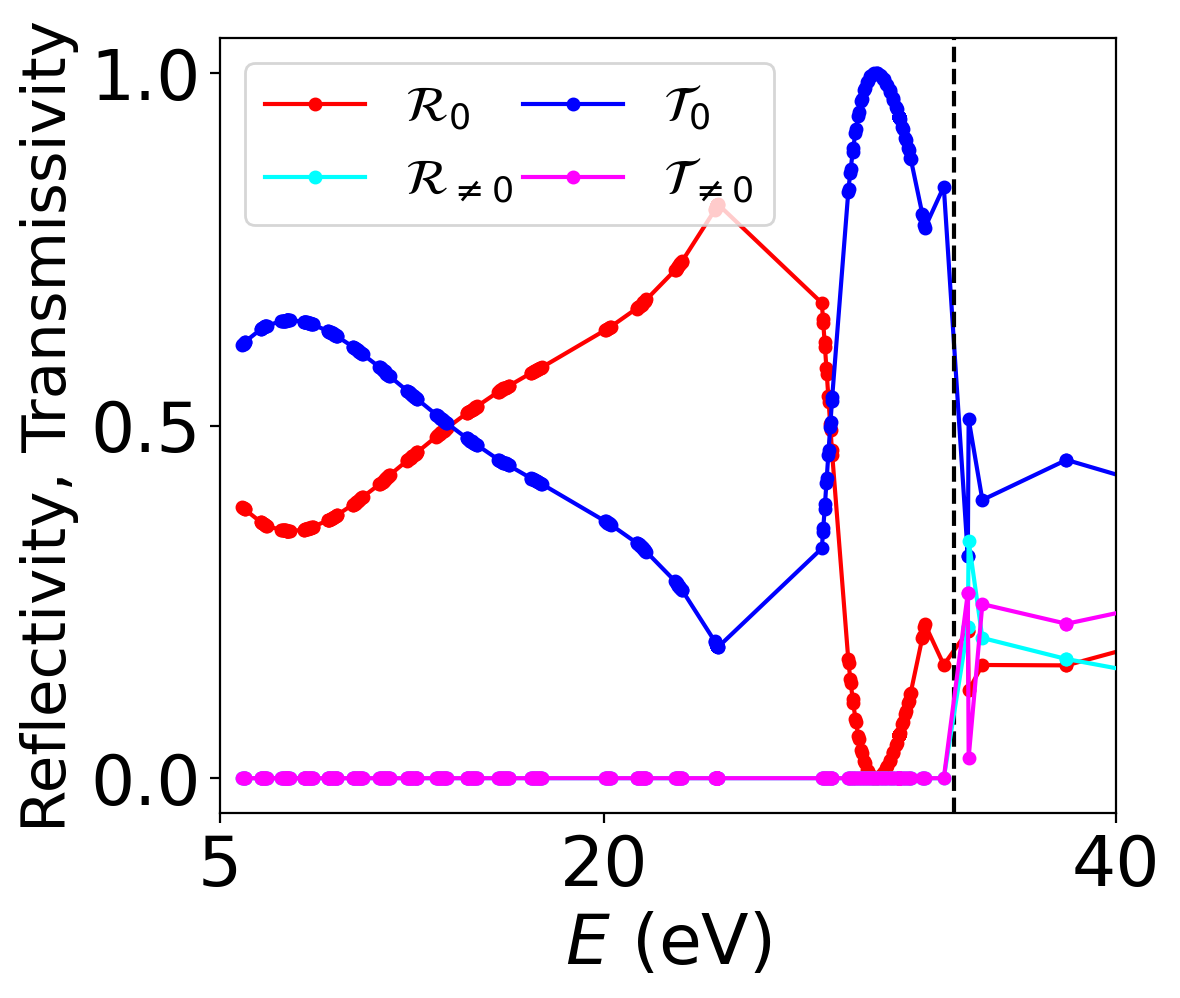}
\end{tabular}
\caption{Calculated LEED reflectivity and transmissivity as a function of
  energy~$E$ for supercell sizes $c$ = 20 {\AA} (a) and $c$ = 40 {\AA}.
  Normal (0) and umklapp ($\ne 0$) components are plotted separately.}
  \label{figure:TR}
  \end{figure}

\begin{figure}[htb]
   \centering
  \begin{tabular}{lll}
    (a)&&(b)\\
    \includegraphics[bb= 0 0 309 258, width=0.3\linewidth]{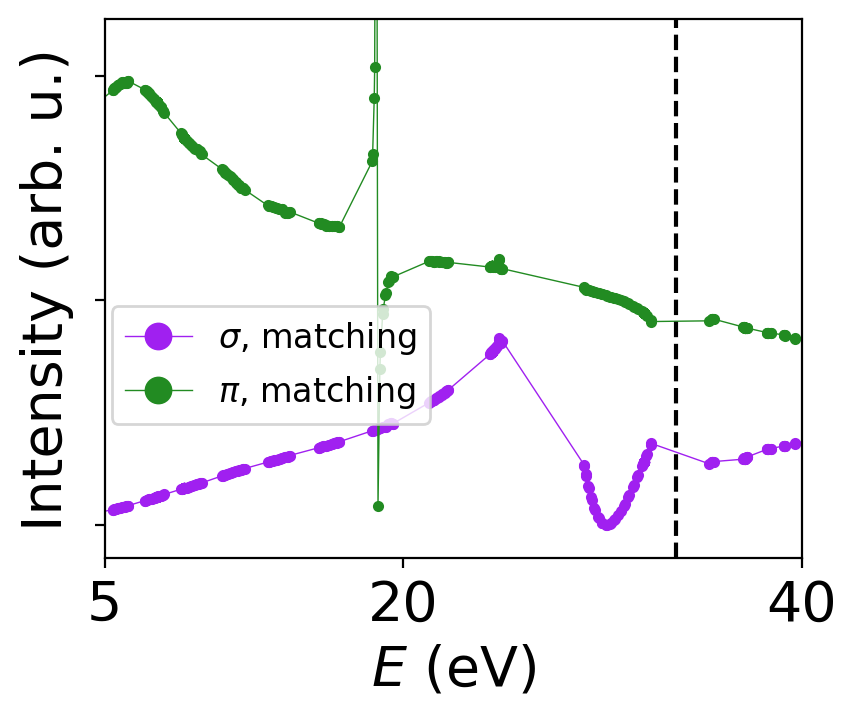}
    && \includegraphics[bb= 0 0 309 258, width=0.3\linewidth]{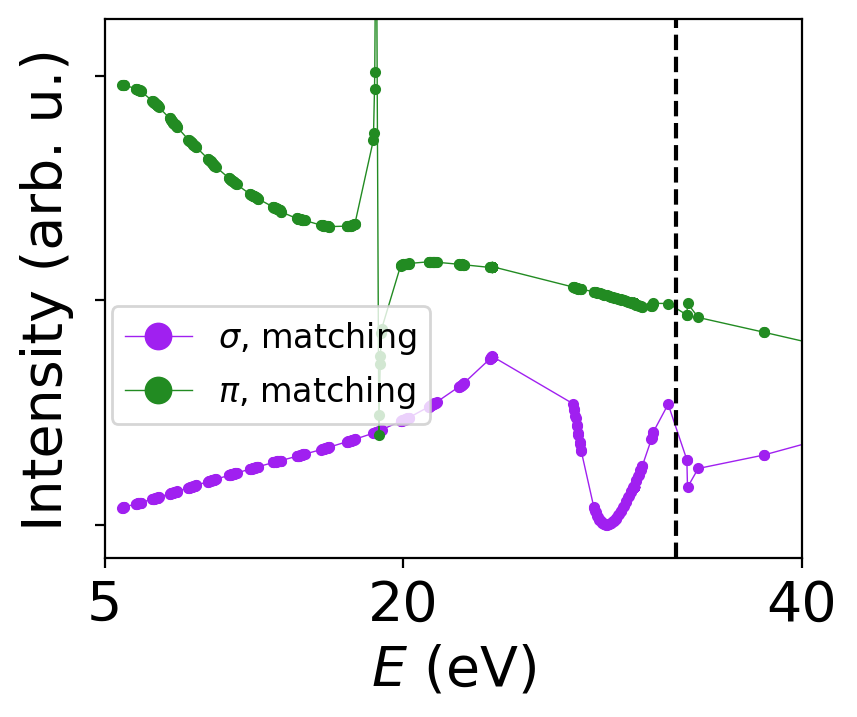}
  \end{tabular}
    \caption{
      Calculated photoelectron intensity obtained with
      $c$ = 20 {\AA} (a) and with $c$ = 40 {\AA} (b).}
  \label{figure:photoint}
    \end{figure}

Here we compare results for supercell size $c=20$~\AA, used in
the main text, with results obtained with a twice larger supercell $c=40$~\AA.
The reflectivity/transmissivity of the LEED state and
photoelectron intensity are shown in Fig. \ref{figure:TR} and Fig. \ref{figure:photoint},
respectively. Panels (a) are for $c$ = 20 {\AA} and panels (b) for $c=40$~\AA.

\clearpage
\section{Wavefunction character of band states at resonance}
\begin{figure}[htbp] 
\begin{tabular}{lll}
\centering
(a)&(b)&(c)\\
\includegraphics[bb=0 0 1271 1120, width=0.3\linewidth ]{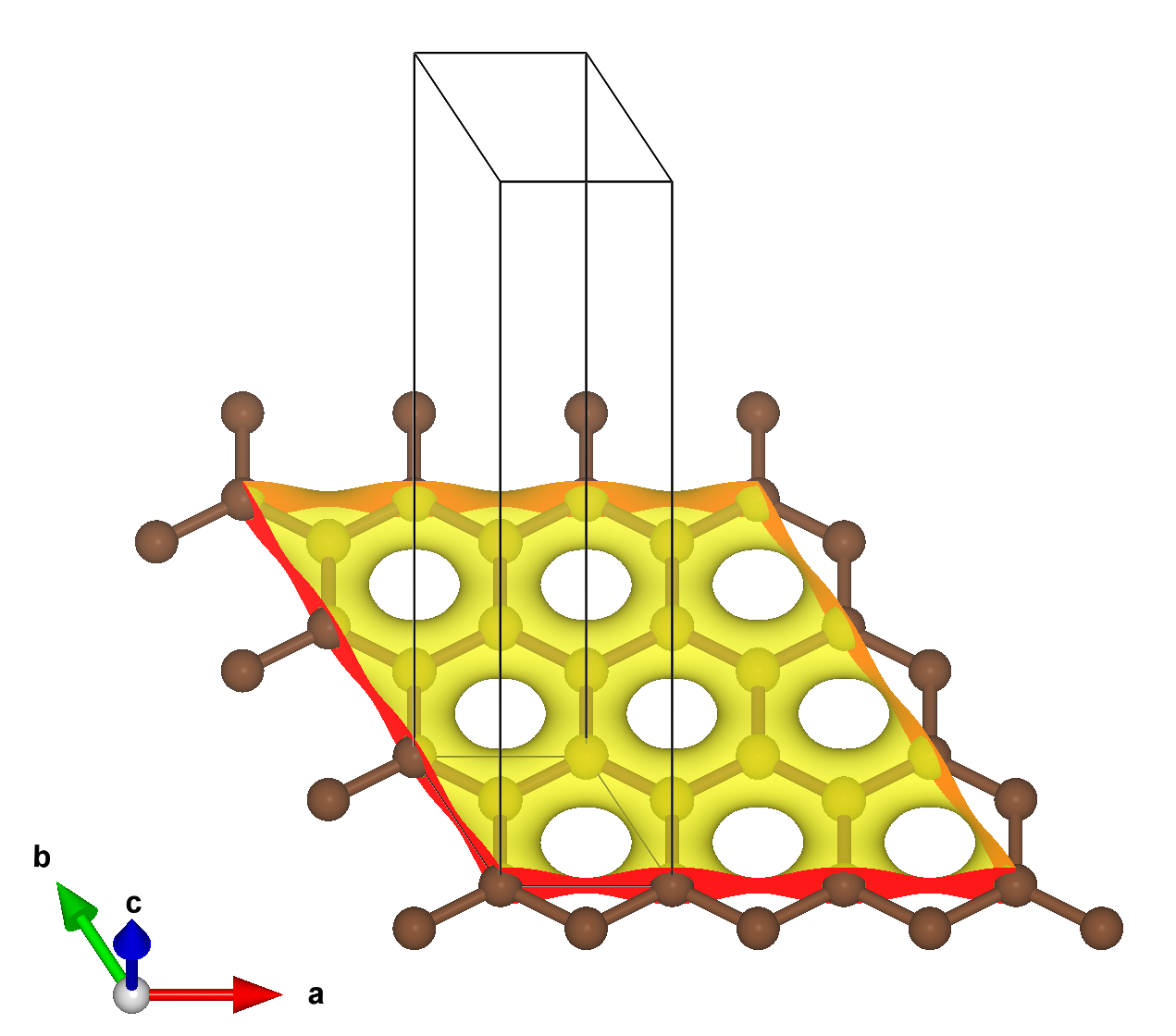} 
&
\includegraphics[bb=0 0 1271 1120, width=0.3\linewidth ]{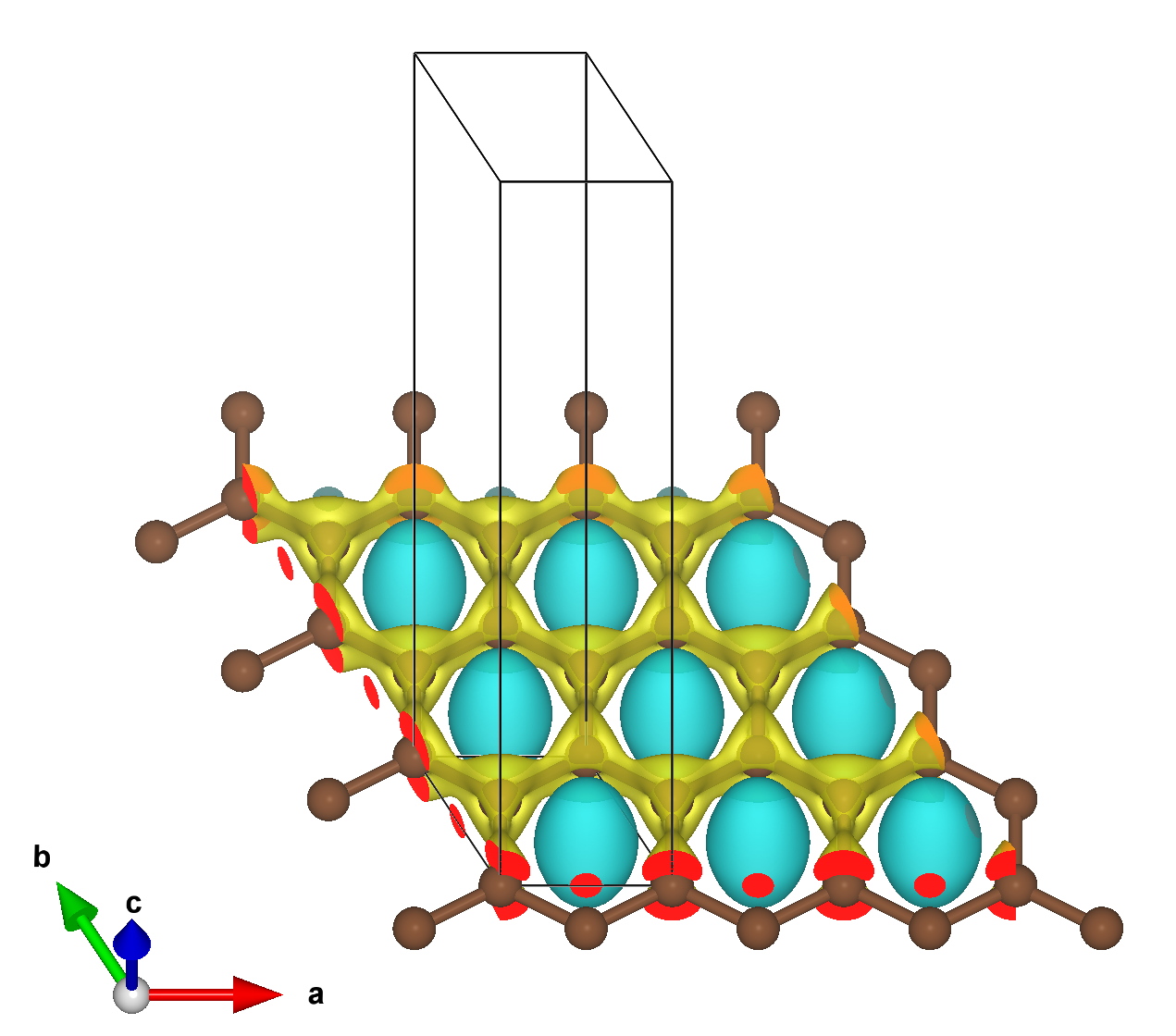}
&
\includegraphics[bb=0 0 460 261, width=0.3\linewidth ]{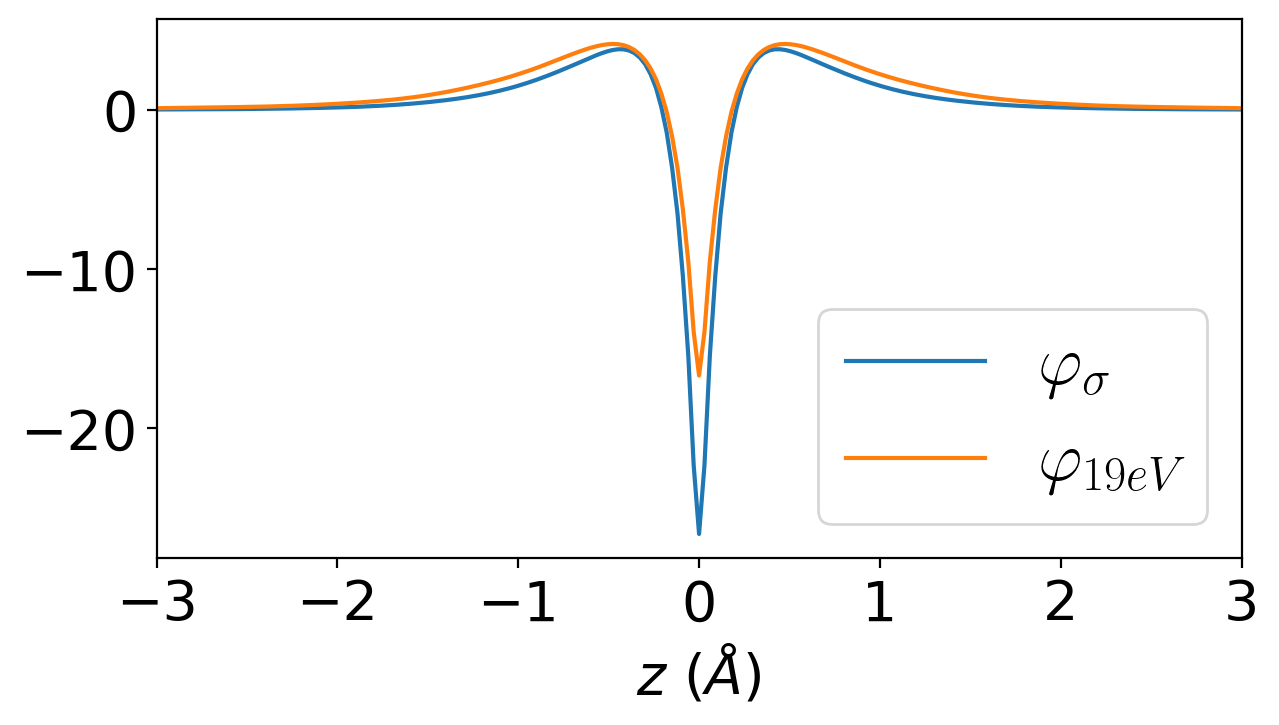}
\end{tabular}
\caption{(a) Isosurface of $\varphi_{\sigma}$. (b) Isosurface of $\varphi_{\rm res}$
  at $E=19$~eV.
  (c) Line plot of $\varphi_{\rm res}$ 
  and $\varphi_{\sigma}$ along $z$-direction at $(x,y)$ = $(0, 0)$ [{\AA}].
  All plots are all-electron wave functions. 
In (a) and (b), the isosurfaces are visualized using the VESTA software \cite{VESTA}.  }
\label{figure:reso}
\end{figure}

Figure \ref{figure:reso}(b) shows the spatial distribution of the wave function
$\varphi_{\rm res}$ at the $\pi$-band resonance, $E=19$~eV 
(state highlighted with blue in Fig. 2 of the main text). 
The symmetry of $\varphi_{\rm res}$ is the same as the $\sigma$ initial state
$\varphi_{\sigma}$ [Fig. \ref{figure:reso}(a)].
In addition, the line plot of $\varphi_{\rm res}$ along $z$-direction through a carbon atom is 
very similar to that of $\varphi_{\sigma}$  [Fig. \ref{figure:reso}(c)].

\bibliography{ref}